\documentclass[usenatbib,twocolumn]{aastex63}

\pdfoutput=1
\usepackage[utf8]{inputenc}
\usepackage{amsmath}
\usepackage{amssymb}
\usepackage{tabularx}
\usepackage{multirow}
\usepackage{url}
\usepackage{amsmath}	
\usepackage{amssymb}	
\usepackage{graphicx}
\usepackage{color}
\usepackage{natbib}
\usepackage{hyperref}
\usepackage{times,txfonts}
\usepackage{bm}
\usepackage[T1]{fontenc}
\usepackage{ae,aecompl}

\bibpunct{(}{)}{;}{a}{}{,} 

\DeclareUnicodeCharacter{00A0}{ }

\usepackage{listingsutf8}
\newcommand{\Ten}[2]{\ensuremath{#1 \times 10^{#2}} }

\newcommand{\unitspace}{\,}

\newcommand{\g}{\ensuremath{\unitspace \mathrm{g}}}

\newcommand{\ud}{\mathrm{d}} % differential d
\renewcommand{\vec}[1]{\ensuremath{\bm{\mathit{#1}}}}

\newcommand{\sigmat}{\sigma_{\mathrm{T}}}

\newcommand{\gammat}{\gamma_{\mathrm{th}}}

\newcommand{\gc}{\gamma_\mathrm{cool}}
\newcommand{\gth}{\gamma_\mathrm{th}}
\newcommand{\ginj}{\gamma_\mathrm{inj}}

\newcommand{\tinj}{t_\mathrm{inj}}
\newcommand{\tacc}{t_\mathrm{acc}}

\newcommand{\tc}{t_\mathrm{cool}}

\newcommand{\A}{{\mathcal A}}
\newcommand{\tf}{t_{\rm flare}}

\newcommand{\omp}{\omega_\mathrm{p}}
\newcommand{\de}{\,c/\omega_\mathrm{p}}
\newcommand{\rL}{\,c/\omega_B}

\newcommand{\ene}[2]{\mathcal{E}_{#1}^{#2}} 

\newcommand{\eneF}{\ene{\mathrm{F}}{}} 
\newcommand{\eneFz}{\ene{\mathrm{F}}{0}}

\newcommand{\eneBperp}{\ene{B}{\perp}} 
\newcommand{\eneE}{\ene{E}{}} 
\newcommand{\eneP}{\ene{\mathrm{prtcl}}{}}
\newcommand{\enePth}{\ene{}{\mathrm{th}}}
\newcommand{\enePnth}{\ene{}{\mathrm{nth}}}
\newcommand{\eneR}{\ene{\mathrm{rad}}{}}

\newcommand{\AB}[1]{#1}

\received{MMM D, YYYY}
\revised{MMM D, YYYY}
\accepted{\today}

\shorttitle{Radiative turbulent flares}
\shortauthors{N\"attil\"a \& Beloborodov}

\begin{document}

\title{Radiative Turbulent Flares in Magnetically Dominated Plasmas}

\author[0000-0002-3226-4575]  {Joonas\,N\"attil\"a}
\affiliation{Physics Department and Columbia Astrophysics Laboratory, Columbia University, 538 West 120th Street, New York, NY 10027, USA}
\affiliation{Center for Computational Astrophysics, Flatiron Institute, 162 Fifth Avenue, New York, NY 10010, USA}
\affiliation{Nordita, KTH Royal Institute of Technology and Stockholm University, Roslagstullsbacken 23, SE-10691 Stockholm, Sweden}
\email{email: jan2174@columbia.edu}

\author{Andrei M.\,Beloborodov}
\affiliation{Physics Department and Columbia Astrophysics Laboratory, Columbia University, 538 West 120th Street, New York, NY 10027, USA}
\affiliation{Max Planck Institute for Astrophysics, Karl-Schwarzschild-Str. 1, D-85741, Garching, Germany}

\begin{abstract}
We perform 2D and 3D kinetic simulations of reconnection-mediated turbulent flares in a magnetized electron-positron plasma, with weak and strong radiative cooling.
Such flares can be generated around neutron stars and accreting black holes.
We focus on the magnetically dominated regime where tension of the background magnetic field lines exceeds the plasma rest-mass density by a factor $\sigma_0 > 1$.
In the simulations, turbulence is excited on a macroscopic scale $l_0$, and we observe that it develops by forming thin, dynamic current sheets on various scales.
The deposited macroscopic energy dissipates by energizing thermal and nonthermal particles.
The particle energy distribution is shaped by impulsive acceleration in reconnecting current sheets, gradual stochastic acceleration, and radiative losses.
We parameterize radiative cooling by the ratio $\A$ of light-crossing time $l_0/c$ to a cooling timescale, and study the effect of increasing $\A$ on the flare. 
When radiative losses are sufficiently weak, $\A<\sigma_0^{-1}$, the produced emission is dominated by stochastically accelerated particles, and the radiative power depends logarithmically on $\A$.
The resulting radiation spectrum of the flare is broad and anisotropic.
In the strong-cooling regime, $\A>\sigma_0^{-1}$, stochastic acceleration is suppressed, while impulsive acceleration in the current sheets continues to operate.
As $\A$ increases further, the emission becomes dominated by thermal particles.
Our simulations offer a new tool to study particle acceleration by turbulence, especially at high energies, where cooling competes with acceleration.
We find that the particle distribution is influenced by strong intermittency of dissipation, and stochastic acceleration cannot be described by a universal diffusion coefficient.
\end{abstract}

\keywords{
Plasma astrophysics (1261);
High-energy astrophysics (739);
Astrophysical magnetism (102);
Computational astronomy (293);
Compact objects (288);
Non-thermal radiation sources (1119);
}

\section{Introduction}

Astrophysical compact objects are often observed as luminous sources of nonthermal radiation. 
Their activity demonstrates efficient dissipation of macroscopic energy stored in magnetic fields and in plasma bulk motions, often in the form of powerful flares. 
These processes energize plasma particles and make them emit broadband radiation spectra, sometimes extending to very high energies.

A possible way for the dissipation to occur is through a macroscopic magnetohydrodynamic (MHD) instability. 
It can excite turbulent motions on a large scale $l_0$ which enable the transfer of energy to small scales, where it can be dissipated.
A special feature of compact objects and their outflows is that the energized plasma can be magnetically dominated.
This condition is often expressed in terms of the magnetization parameter $\sigma=B^2/4\pi\rho c^2$, where $B$ is the magnetic field, and $\rho$ is the plasma mass density. 
The magnetization $\sigma>1$ is likely in the coronae and jets of accreting black holes; 
$\sigma\gg 1$ also occur in pulsar magnetospheres and their winds.

Excitation of turbulent motions in a magnetically dominated plasma with amplitude $\delta B/B\sim 1$ implies that a large energy per particle becomes available for dissipation. 
Furthermore, turbulence can stochastically accelerate a fraction of particles to extremely high energies (e.g. \citealt{Petrosian2012}).
A complete model of this process must self-consistently follow the plasma waves and individual particle dynamics, which can be done with advanced numerical simulations.

A special feature of the magnetically dominated plasma is that its Alfv\'en speed is relativistic, $V_{\mathrm{A}} = \sqrt{\sigma/(1+\sigma)} \approx c$.
Relativistic kinetic turbulence simulations have only recently become feasible thanks to the increase in computational resources \citep{Zhdankin_2017b, Zhdankin_2017a, Zhdankin_2019, Comisso_2018, Comisso_2019, Nattila_2019, Wong_2019}.
In agreement with theoretical expectations, the numerical experiments demonstrated that a fraction of plasma particles experience stochastic acceleration to very high energies, until their Larmor radius becomes comparable to $\ell_0$.

In astrophysical objects, turbulent heating of the plasma can be accompanied by significant radiative losses. 
The losses limit particle acceleration and the growth of the plasma temperature. 
Furthermore, radiative losses may affect the development of the turbulence itself.
Most of previous work on radiative turbulence was analytical \citep[e.g.,][]{Thompson_2006, Uzdensky_2018, Sobacchi_2019, Zrake_2019}
The only existing kinetic simulations of radiative relativistic turbulence were recently performed by \citet{Zhdankin_2020}. 
Their simulation setup assumed steady driving of turbulence in a magnetized electron-positron plasma with a strong level of radiative losses, which completely suppressed stochastic particle acceleration.
 
In the present paper, we perform kinetic simulations of turbulent {\it flares}. 
We envision a sudden excitation of turbulence by an MHD instability, which deposits a macroscopic energy comparable to the total magnetic energy of the system.
The initial disturbance on a large scale $l_0$ is followed by the development of turbulent plasma motions and the eventual dissipation of the injected magnetic energy.
We investigate how the deposited turbulence energy is radiated, how the radiative losses affect particle acceleration, and what spectra can be radiated by the turbulent flares.
For simplicity, all our simulations will assume that the plasma is made of electrons and positrons and that the plasma is optically thin, so that the emitted radiation freely escapes. 
The opposite, optically thick, regime was recently investigated in the context of gamma-ray bursts by \cite{Zrake_2019}.

We perform both two-dimensional (2D) and three-dimensional (3D) radiative kinetic simulations to model the flares.
Our setup of initial conditions is similar to that in \citet{Comisso_2018,Nattila_2019, Comisso_2019}.
Remarkably, for this setup the turbulence development and particle acceleration picture in 2D is similar to the results of full 3D simulations \citep{Comisso_2019}. 
We call this turbulence {\it reconnection mediated}, as we observe that the turbulence develops by forming reconnecting current sheets, in contrast to the canonical picture of an Alfv\'en-wave cascade.

2D simulations have lower computational costs and can be performed with particularly long durations and high resolutions. 
We use many 2D simulations to systematically study the effects of radiative losses.  
We also perform a few 
large-scale 3D simulations to test their difference from the 2D models.
All our simulations are performed with the open-source kinetic code \textsc{runko} 
\citep{Nattila_2019}. 

The paper is organized as follows. 
The simulation setup is described in Section~\ref{sect:numerics}. 
Section~\ref{sect:turb} presents our results for turbulent flares without cooling.
Then, in Section~\ref{sect:radturb}, we use analytical estimates to discuss the expected effects of radiative losses and define two cooling regimes: weak and strong. 
The full radiative simulations are presented in Sections~\ref{sect:results} and 
\ref{sect:flares}.
Finally, conclusions are given in Section~\ref{sect:conclusions}.

\section{Simulation Setup}\label{sect:numerics}

\subsection{Pre-flare State}
\label{equilibrium}

The unperturbed equilibrium state is a homogeneous neutral pair plasma with a temperature $T_0$. 
The corresponding dimensionless temperature is 
\begin{equation}
    \theta_0 = \frac{k_{\mathrm{B}} T_0}{m_e c^2},
\end{equation}
where $k_{\mathrm{B}}$ is the Boltzmann constant, $m_e$ is the electron rest mass, and $c$ is the speed of light. 
All models shown in this paper have $\theta_0 = 0.3$, which corresponds to a mean particle Lorentz factor of $\gammat \approx 1+3\theta_0\approx 1.6$. We also performed simulations with $\theta_0$ ranging from $10^{-4}$ up to $0.6$, with similar results.
The choice of $\theta_0$ is  unimportant as long as the initial thermal energy is much smaller than the energy of the injected turbulence.

The pre-flare plasma is magnetized with a uniform magnetic field $\boldsymbol{B}_0$. 
The dimensionless magnetization parameter is defined as
\begin{equation}
    \sigma_0 = \frac{B_0^2}{4\pi \rho_0 c^2},
\end{equation}
where $\rho_0 = n_0 m_e$ is the plasma rest-mass density, and $n_0 = n_- + n_+$ is the number density of electrons and positrons.
The magnetization parameter that takes into account heat contribution to the plasma inertia is given by
\begin{equation}
  \sigma = \frac{\sigma_0}{\gth}\approx \frac{\sigma_0}{1+3\theta_0}.
\end{equation}
In this paper, we focus on the magnetically dominated regime of $\sigma_0 > 1$, and our fiducial simulation setup has $\sigma_0 \approx 16$ and $\sigma = 10$.
In addition, we performed simulations with $\sigma_0 \approx 2$, $8$, $30$ (and $\sigma = 1$, $5$, $20$).

The magnetized plasma is described by two characteristic frequencies: 
the plasma frequency $\omp$ and the frequency of Larmor rotation $\omega_B$. 
They are given by 
\begin{equation}
    \omp =\left(\frac{4\pi e^2 n_0}{m_e}\right)^{1/2}, 
    \qquad  
    \omega_B =\frac{e B_0}{m_e c},
\end{equation}
where $e$ is the electron charge. 
Note that $\omega_B/\omp=\sigma_0^{1/2}$.
The two frequencies define two characteristic scales of the problem: 
the plasma skin depth $c/\omp$ and the Larmor radius of nonrelativistic particles $c/\omega_B$.
The plasma is strongly magnetized in the sense that $\rL$ is  smaller than $c/\omega_p$, which in turn is much smaller than the size of the system. When the plasma is heated to $\gth\gg 1$, the average electron inertial mass increases to $\gth m_e$, the effective plasma frequency becomes $\omega_p/\sqrt{\gth}$, and the typical Larmor frequency of thermal electrons becomes $\omega_B/\gth$.
% For relativistic particle motions these become $\omega_p/\sqrt{\gamma}$ and $\omega_B/\gamma$.

\subsection{Exciting the Turbulent Flare}
\label{excitation}

Let us choose the $z$-axis along the unperturbed magnetic field $\vec{B}_0$. 
In 2D systems all perturbed quantities will remain independent of $z$ and dynamics will occur in the $x$-$y$ plane.
In 3D simulations we additionally perturb the system in the $z$ direction.

Turbulence is created by starting from a nonequilibrium, excited state in the same way as in \citet{Comisso_2018,Nattila_2019, Comisso_2019}.
The plasma is initially at rest and carries no electric current;
it has a uniform density $n_0$ and temperature $\theta_0$. 
The initial excited state differs from the equilibrium state described in Section~\ref{equilibrium} only by the presence of an additional magnetic field perpendicular to $\vec{B}_0$: $\vec{B}_{\perp}=(B_x,B_y)$.
This field is described by its Fourier components as follows:
\begin{align}
    B_x &= \phantom{+}\sum_{l,m,n} \beta_{lm} m \sin(k_l x+\phi_{lmn})\cos(k_m y+\psi_{lmn})\sin(k_n z + \chi_{lmn})  \label{eq:B_perturb1}, \\
    B_y &= -          \sum_{l,m,n} \beta_{lm} l \cos(k_l x+\phi_{lmn})\sin(k_m y+\psi_{lmn})\sin(k_n z + \chi_{lmn}),  \label{eq:B_perturb2}
\end{align}
where $l,m \in \{1,\ldots,N_\perp\}$ are the perpendicular mode numbers, $n \in \{1,\ldots,N_\parallel\}$ is the parallel (along $\vec{\hat{z}}$) mode number, $k_l = 2\pi l/L$, $k_m = 2\pi m/L$, and $k_n = 2\pi n/L$ are the wavenumbers along $x$, $y$ and $z$, respectively, and $\phi_{lmn}$, $\psi_{lmn}$, $\chi_{lmn}$ are random phases.
The parameters
\begin{equation}
    \beta_{lm} \equiv 
        %\frac{1}{2\sqrt{2}} 
        %\frac{1}{N_\perp \sqrt{N_\parallel}}
        \frac{\sqrt{2}}{\sqrt{N_\parallel}}
        \frac{2}{N_\perp}
        \frac{B_{\perp}^{\mathrm{rms}} }{\sqrt{l^2 + m^2}}
\end{equation}
set the amplitude of the perturbations $\delta B=B_{\perp}$, which is described by the rms value $B_\perp^{\mathrm{rms}} = \sqrt{\langle B_\perp^2 \rangle}$, where $\langle \ldots \rangle$ denotes volume average over the simulation domain.
For 2D configurations the normalization reduces to $\beta_{lm} = 2 B_\perp^{\mathrm{rms}}/(N_\perp \sqrt{l^2 + m^2})$ and $\sin(k_n z + \chi_{lmn}) \rightarrow 1$ in Equations~\eqref{eq:B_perturb1} and \eqref{eq:B_perturb2}.
The perpendicular field satisfies $\langle \vec{B}_\perp \rangle = 0$.
The perturbation is non-helical.

In our fiducial 2D setup we use $N_\perp=8$ modes, which results in stirring turbulence on scale $l_0 \approx 125 \de$. 
We also made test runs with $N_\perp=4$ and $16$. 
We chose $N_\perp=8$ because in this case $l_0$ is sufficiently small to excite many turbulent eddies in the box, and sufficiently large to be far from the microscopic plasma scale. 
In 3D simulations we are limited by the computational cost and therefore forced to select a smaller number of modes;
we use $N_\perp=3$ which corresponds to a similar $l_0 \approx 140 \de$.
Additionally, we perturb the system in the $z$ direction with two sinusoidal modes, $N_\parallel = 2$.

Our fiducial model has the initial perturbation amplitude $B_{\perp}^{\mathrm{rms}}/B_0 = 1$. 
This setup is designed so that one can think of scale $l_0$ as the size of turbulent eddies for which $\delta B/B\sim 1$. 
Note that this way of triggering turbulence is quite violent. 
The initial state is far out of pressure balance leading to a quick rearrangement of the system and thus exciting mildly relativistic plasma motions in the $x$-$y$ plane.
The initial rearrangement causes transient phenomena on the timescale $\sim l_0/c$, before relaxation into a quasi-steady turbulent state.
We focus on the latter, quasi-steady stage, which lasts a much longer time of tens of $l_0/c$.
Since there is no driving of turbulence apart from the strong initial perturbation described above, the turbulent motions eventually dissipate. 
We follow this process to a time $t$ of at least $20 \,l_0/c$. 
The durations of some simulations were extended to $t \sim 100 \,l_0/c$.

An alternative way of exciting similar turbulence is by perturbing the $J_z$ current with an oscillating Langevin antenna \citep{TenBarge2014} instead of perturbing $\vec{B}$. 
We have verified that this leads to similar results.

\subsection{Numerical Implementation}

Fully kinetic calculations are required to study how turbulence energizes plasma particles. 
We use relativistic particle-in-cell (PIC) simulations, where the field evolution is calculated on a grid, and the plasma is represented by a large number of charges moving through the grid and creating electric currents. 
All our simulations are performed with the recently developed code \textsc{Runko}%
\footnote{Public release \textsc{v2.0}: Orange (commit 97202eb).}
\citep{Nattila_2019}, designed as a modern, massively parallel, C++14/Python3 platform for plasma simulations.
The PIC module in \textsc{Runko} uses a second order finite-difference time-domain electromagnetic field solver, a charge-conserving current deposition scheme, and digital current filtering.
Particles are propagated in time with a relativistic Boris pusher \citep[see][for details]{Nattila_2019}.
For the present work, we have modified the code to include radiative losses of particles as 
described in Sect.~\ref{sect:cooling_model}.

The code evolves all three ($x$, $y$, $z$) components of the fields and the particles' velocities. 
Periodic boundary conditions are imposed on the computational box. 
At each time step we perform $8$ digital current filtering passes (with a $3$-point binomial filter) to damp out unphysical high-frequency numerical noise.

In our 2D simulations, the domain is a square in the $x$-$y$ plane of size $L = 1024\de$. 
The square is covered by a Cartesian grid of size $5120^2$, so that the (nonrelativistic) plasma skin depth $\de$ is resolved with $5$ grid cells.  
The plasma is simulated with $32$ particles per cell per species.  
We have benchmarked the validity of this setup against shorter simulations with up to $256$ particles per cell per species, $10$ grid points per skin depth, and no current filtering. 
The results were found to be well converged.  We have also tested different computational box sizes from $L \omp/c \approx 100$ up to $\sim 7000$ (corresponding to a maximum grid size of $20480^2$). 
We found that a minimum scale separation of $l_0 \omp/c \sim 100$ is needed to properly capture the phenomena described below in this paper. 

In our 3D simulations, the domain is a cube of side $L = 426 \,\de$ covered by $1280^3$ grid cells.
We resolve the (nonrelativistic) plasma skin depth with $3$ cells and use $2$ particles per cell per species to model the plasma. 
Four current filtering passes are performed on each time step.
Similar to the 2D case, the validity of these simulation parameters was benchmarked against shorter simulations with a maximum size of $L = 640 \,\de$ (corresponding to a resolution of $1920^3$).

\section{Relativistic Kinetic Turbulence}\label{sect:turb}

\subsection{The Role of Magnetic Reconnection}

\begin{figure*}
\centering
    \includegraphics[trim={0.0cm 0.1cm 0.0cm 0.0cm}, clip=true, width=0.47\textwidth]{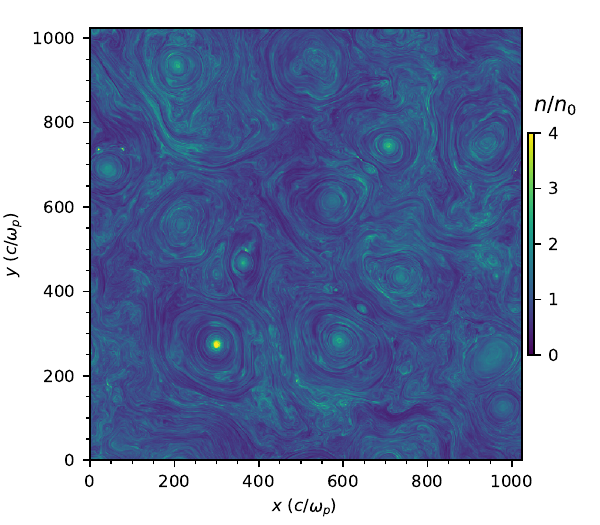}
    \includegraphics[trim={0.5cm 0.0cm 0.8cm 0.5cm}, clip=true, width=0.35\textwidth]{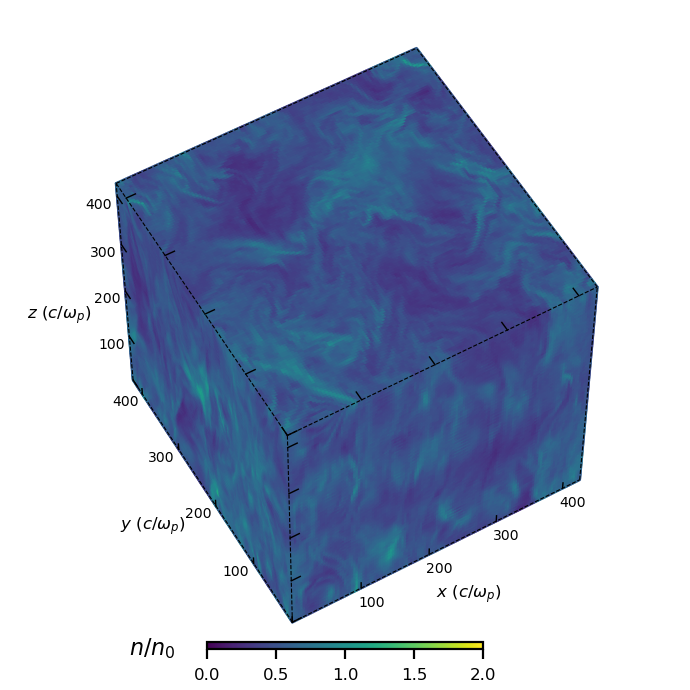}
    \includegraphics[trim={0.0cm 0.1cm 0.0cm 0.4cm}, clip=true, width=0.47\textwidth]{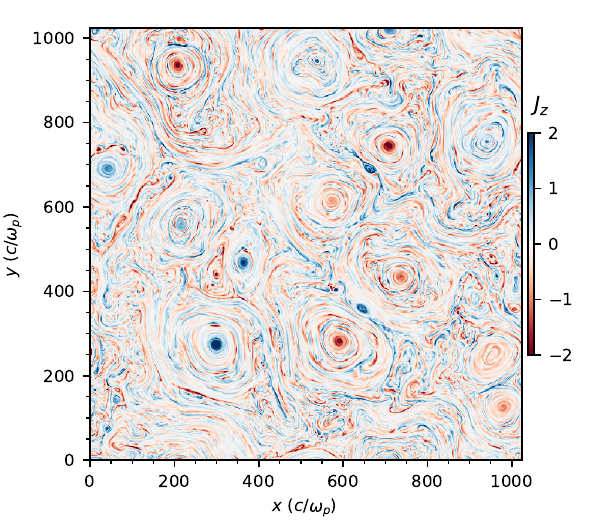}
    \includegraphics[trim={0.5cm 0.0cm 0.8cm 0.5cm}, clip=true, width=0.35\textwidth]{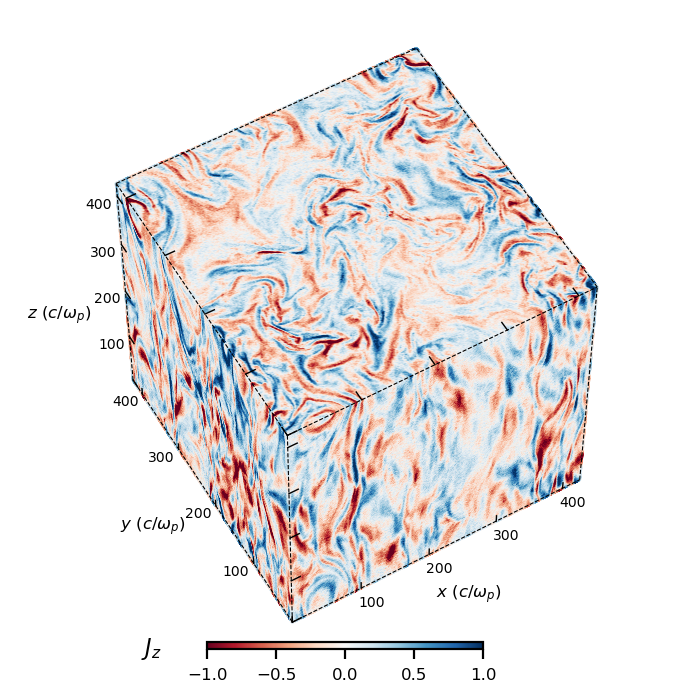}
    \includegraphics[trim={0.0cm 0.0cm 0.0cm 0.4cm}, clip=true, width=0.47\textwidth]{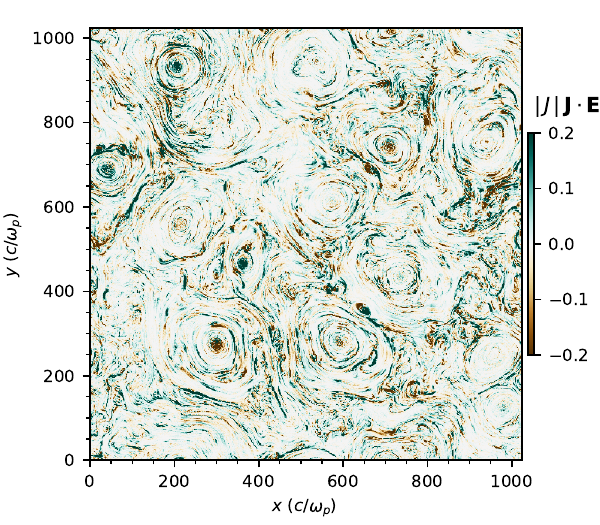}
    \includegraphics[trim={0.5cm 0.0cm 0.8cm 0.5cm}, clip=true, width=0.35\textwidth]{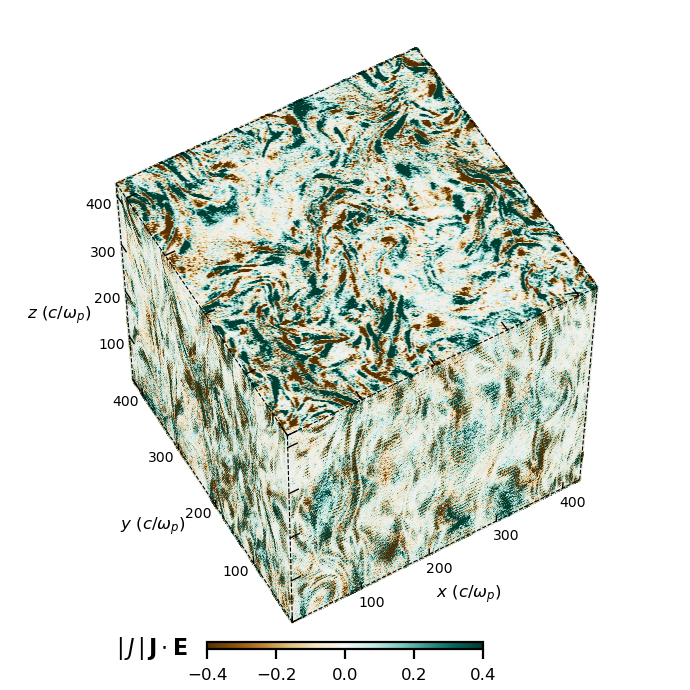}
\caption{\label{fig:visuals}
    General appearance of the relativistic turbulent plasma for the fiducial $\sigma_0 = 16$ magnetization.
    Left panels show the full 2D simulation domain at $t \approx 10\,l_0/c$ and right panels show the periphery of the 3D simulation box at $t \approx 5\,l_0/c$.
    The dominant background magnetic field ${\mathbf B}_0$ is oriented along the $\vec{z}$-axis (out of the plane in the 2D figures).  
    Top row shows the plasma density $n/n_0$ (in units of the initial plasma density $n_0$), 
    middle row shows current density $J_z$ (in units of $e n_0 c$), 
    and bottom row shows the local dissipation rate coefficient ${\cal D}_J$, which is proportional to $|J|\,\vec{E}\cdot\vec{J}$ and defined in Equation~(\ref{eq:diss}).
    Magnetic flux ropes appear as round overdense structures in the 2D setup.
    Current sheets are constantly being created at the interfaces of the colliding and merging flux ropes.  
    They are sites of strong, localized, intermittent dissipation as observed in the bottom panels.
}
\end{figure*}

\begin{figure}[t]
\centering
    \includegraphics[width=0.47\textwidth]{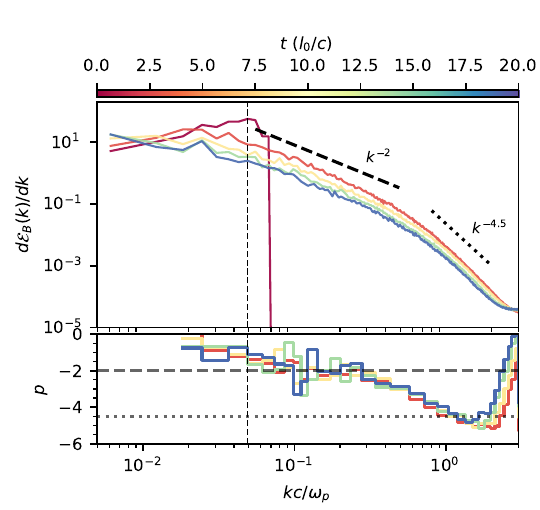}
\caption{\label{fig:spec}
    Magnetic field energy spectrum $d\ene{B}{} (k)/dk$ for the fiducial 2D simulation with $\sigma_0 = 16$.  
    Different colors correspond to different times $t$ as indicated in the color bar. 
    Bottom panel shows the spectral slope computed in a narrow moving time window.
    For comparison, two slopes are indicated: $-2$ (dashed line) and $-4.5$ (dotted line).
    Vertical dashed line shows the location of the injection scale, $l_0$.
}
\end{figure}

\begin{figure}[t]
\centering
    \includegraphics[width=0.47\textwidth]{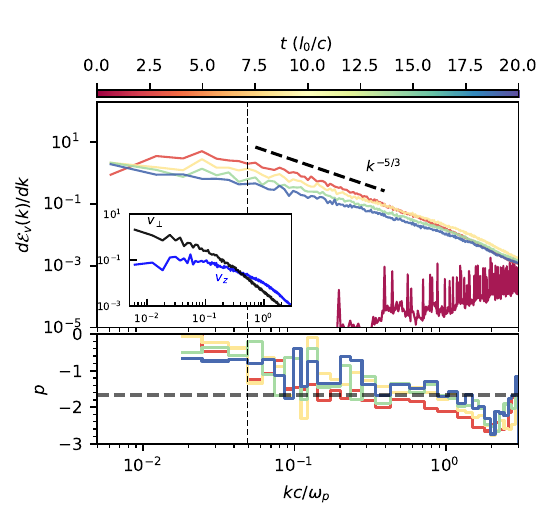}
\caption{\label{fig:spec_kin}
    Spectrum of the plasma bulk motions, $d{\cal E}_v/dk$, for the fiducial 2D simulation with $\sigma_0 = 16$.
    Different colors correspond to different times $t$, similar to Figure~\ref{fig:spec}. 
    Bottom panel shows the spectrum slope computed in a narrow moving time window.
    For comparison, a spectral slope of $p = -5/3$ (dashed line) is shown.
    The inset shows the separated energy spectra of the perpendicular velocity component $\propto v_\perp^2$ (black curve) and out-of-the plane component $\propto v_z^2$ (blue curve) at $t \approx 10~l_0/c$.
}
\end{figure}

Turbulence in a magnetized plasma is usually expected to develop  an energy cascade toward small spatial scales $l\ll l_0$, in both 2D and 3D models \citep[e.g.,][]{Biskamp_2003}.
MHD cascades, involving fluctuations of the magnetic field, $\delta\vec{B}$, and the plasma velocity, $\delta\vec{v}$,  transport the energy mainly perpendicular to the background magnetic field, towards larger perpendicular wavenumbers, $k_\perp = 2\pi/l_\perp$ \citep{Goldreich_1995}.
A self-similar cascade has a power-law distribution of fluctuation amplitudes as a function of $k_\perp$,
\begin{equation}
    \delta v(k_\perp) \sim \delta B(k_\perp) \propto k_\perp^q,
\end{equation}
corresponding to a magnetic energy spectra (energy per wavenumber interval) 
\begin{equation}
    \frac{d \ene{B}{} }{d k} 
    \propto \frac{ \delta \vec{B}^2(k_\perp) }{k_\perp} 
    \propto k_\perp^p 
    = k_\perp^{2q -1}.
\end{equation}
The canonical Kolmogorov slope of $p=-5/3$ corresponds to $q = -1/3$. 
Steeper slopes were also found, in particular in the force-free limit \citep{Li_2019}. 
The cascade extends down to scale $l_\nu$ at which dissipative processes stop the nonlinear transfer of turbulence energy to smaller scales.

In magnetically dominated plasma the turbulent motions are mainly shear motions perpendicular  to the background field $\vec{B}_0$. 
The turbulence is nearly transverse ($\vec{v}\perp\vec{B}_0$) and has a small amplitude $\delta B\ll B_0$ on small scales $l$, i.e.  it weakly bends the guide field lines. 
The transverse plasma motions follow the displacements of the magnetic field lines as long as the ideal MHD description holds.
Their decoupling occurs on small kinetic scales and leads to dissipation.

Some previous studies of turbulence emphasized the dissipative role of magnetic reconnection, on both large and small scales \citep[e.g.,][]{Biskamp_2003, Loureiro_2020}.
The scale of this process depends on how the turbulence is generated.
Magnetic reconnection is quickly activated in our simulations, and strongly affects the development of plasma motions on small scales.
A typical state of the computational box after the development of turbulence is shown in Figure~\ref{fig:visuals}.
We observe that the flux ropes twisted by the initial perturbations have quickly developed thin current sheets between them, which begin to reconnect, generating smaller flux ropes. 
At the same time, there is an inverse process of \textit{coagulation} of flux ropes. 
The system remains highly dynamic as the flux ropes move around, collide, and merge.
Flux tubes with the same current polarity attract, while opposite polarities repel each other.

In this reconnection-mediated turbulence, current sheets serve as the sites for  turbulence dissipation. We observe that the magnetic field fluctuations
efficiently decay via reconnection in the numerous current sheets on various scales. 
Note that, in contrast to MHD models, kinetic plasma simulations do not need any prescriptions for resistivity. 
Instead, they follow the development of the tearing instability of the current sheets and the resulting dissipation from first principles.
At large scales, $l \omega_p/c \rightarrow\infty $, kinetic simulations reproduce the MHD model.%
\footnote{%
A comparable MHD simulation that captures the reconnection-mediated dissipation regime needs to have a very large magnetic Reynolds number of $Rm = v L/\eta \gtrsim 10^4$ (where $\eta$ is the resistivity) in order to model the sheet tearing correctly. 
This translates to very large simulation box sizes that are only starting to be probed by contemporary simulations \citep{Dong_2018}.
}

Figure~\ref{fig:spec} shows the evolution of the turbulence magnetic energy spectrum in our fiducial 2D simulation with $\sigma_0 \approx 16$. The full 2D spectrum in the $\vec{k}=(k_x,k_y)$ plane is defined by $d\mathcal{E}_B/d\vec{k} = \frac{1}{8\pi}|\vec{\hat{B}}(\vec{k})|^2$, where $\vec{\hat{B}}$ is the Fourier transform of  $\vec{B}=(B_x,B_y,B_z)$; 
Figure~\ref{fig:spec} shows $d\mathcal{E}_B/dk$ obtained by integrating $d\mathcal{E}_B/d\vec{k}$ over annuli $2\pi k\,\delta k$.
The initial perturbations occupy the region  $k\de<0.07$, and the main injection scale $l_0$ corresponds to $k_0\de\approx 0.05$. 
The relaxation to a fully developed spectrum occurs in about one dynamical time $l_0/c$. 
The spectrum quickly approaches a power-law form with a slope of $p \approx -1.9 \pm 0.4$ between $k_0$ and $0.7 k_c$, where
\begin{equation}
 k_c=\frac{2\pi c}{\omega_{\rm p}},
\end{equation}
and the ideal MHD approximation breaks. We also studied the turbulence spectrum in the 3D simulation, and found similar results when viewed in the $\vec{k}_\perp$-space. 
 
Next, we have analyzed the plasma bulk motions.
The kinetic energy density of the bulk motions in the nonrelativistic approximation is $\rho v^2/2$, where $\rho$ is the plasma rest-mass density, and $\vec{v}$ is the local bulk velocity, which  is found by averaging particle velocities (both species) in each computational cell. 
The kinetic energy spectrum was evaluated as $d\mathcal{E}_v/d\vec{k} = \frac{1}{2} |\hat{\vec{w}}(\vec{k})|^2$, where $\hat{\vec{w}}$ is the Fourier transform of $\vec{w} = \sqrt{\rho} \vec{v}$ \citep[see e.g.,][]{Kritsuk_2007}.
Then, $d\mathcal{E}_v/dk$ was obtained by integrating $d\mathcal{E}_v/d\vec{k}$ over annuli $2\pi k\,\delta k$ (in a 2D simulation $\vec{k}$ is in the $x$-$y$ plane).
The evolution of the bulk motion spectrum is shown in Figure~\ref{fig:spec_kin} for our fiducial 2D simulation with $\sigma_0 \approx 16$.
At time $ t > 3~l_0/c$, we find that $d\mathcal{E}_v/dk$
has a slope of $p_v = -1.5 \pm 0.3$ in the inertial range $k_0<k< 0.7 k_c$. 
It is somewhat shallower than the magnetic energy spectrum. 

In both 2D and 3D simulations, we observed that the kinetic power at small scales $k\gtrsim 0.7 k_c$ originates from plasma motions along the $z$ direction, i.e. along the background field $\vec{B}_0$. 
This is in contrast to motions on large scales, which are dominated by transverse velocities. 
The change from mainly transverse to mainly $z$ motions approximately coincides with the spectral break in the magnetic field spectrum, where its slope steepens to $p \approx -4.5 \pm 0.4$.

\begin{figure*}
\centering
    \includegraphics[trim={0.0cm 0.1cm 0.0cm 0.0cm}, clip=true, width=0.47\textwidth]{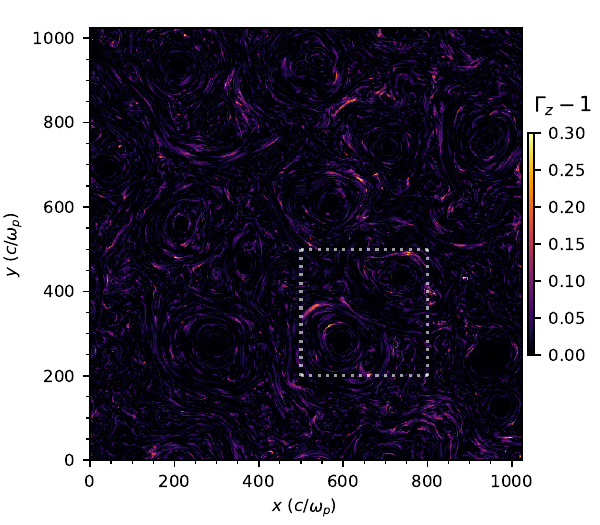}
    \includegraphics[trim={0.0cm 0.1cm 0.0cm 0.0cm}, clip=true, width=0.47\textwidth]{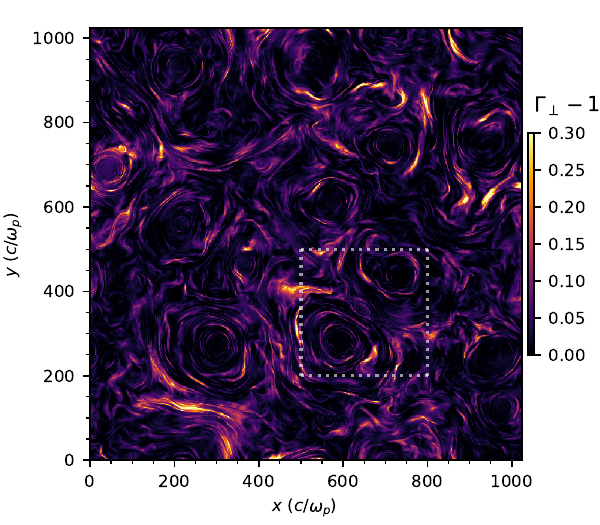}
    \includegraphics[trim={0.0cm 0.1cm 0.0cm 0.0cm}, clip=true, width=0.47\textwidth]{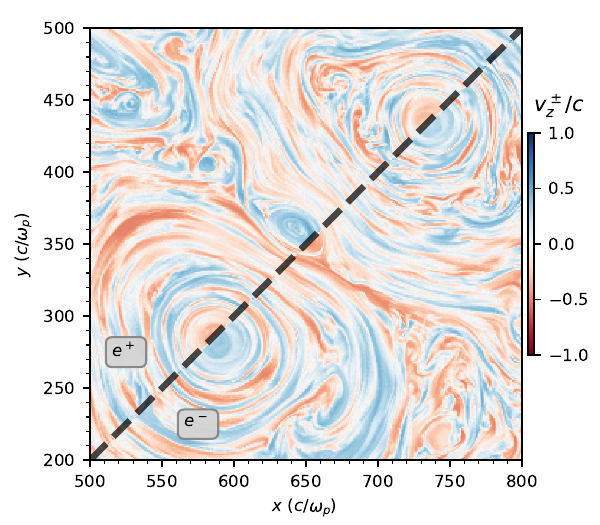}
    \includegraphics[trim={0.0cm 0.1cm 0.0cm 0.0cm}, clip=true, width=0.47\textwidth]{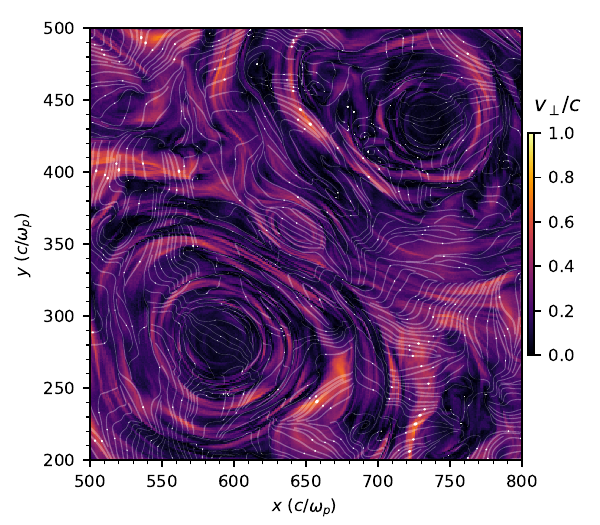}
\caption{\label{fig:vel_visuals}
    Visualization of the plasma motions in the fiducial 2D simulation with magnetization of $\sigma_0 = 16$ at $t \approx 10\,l_0/c$.
    Top-left panel shows a map of $\Gamma_z -1$, where $\Gamma_z=(1-v_z^2/c^2)^{-1/2}$ is the Lorentz factor of the plasma velocity in the $z$ direction, $v_z$ (out-of-plane velocity component).
    Top-right panel shows a similar quantity $\Gamma_\perp -1$ (where $\Gamma_\perp = (1-v_\perp^2/c^2)^{-1/2}$) for the bulk motion $\vec{v}_\perp$ in the $xy$ plane.
    White dashed rectangles indicate the location of the zoom-in regions shown in the bottom panels, where one can see strong bulk flows at small scales.
    Bottom panels also visualize the direction of these flows: 
bottom-right panel shows $\vec{v}_{\perp}$ with the plasma streamlines (white curves with arrows), and bottom-left panel shows $v_z^\pm$ for electrons ($v_z^-$; lower-right corner) and positrons ($v_z^+$; upper-left corner).
    The $\vec{v}_\perp$ motions reflect the active magnetic reconnection --- fast motions along the reconnection layers and the inflows feeding plasma into the layers.
    The $z$ motions of $e^\pm$ are opposite in sign and relativistic at the locations of the current sheets, $v_z^\pm \sim c$, indicating that the sheets' thickness is regulated by charge starvation.
}
\end{figure*}

The plasma motions are visualized in Figure~\ref{fig:vel_visuals}. 
We observe the following:
\begin{enumerate}
    \item
 The bulk velocities of the electron and positron fluids along $z$ have opposite directions and almost equal magnitudes, $v_z^+\approx -v_z^-$, so the net $v_z$ is negligible across the $k$-spectrum except small kinetic scales where deviations from charge neutrality occur and $J_z$ becomes accompanied by $v_z$ fluctuations (exceeding the small-scale $v_\perp$).
    \item
 The currents, and hence $v_z^\pm$, are concentrated in thin current sheets. 
Note that $v_z^\pm$ are comparable to $c$, i.e. the current sheets are not far from being charge starved. 
The initial large-scale perturbations are smooth and drive the characteristic current $j_0=(c/4\pi)B_0/l_0$.
The ratio 
\begin{equation}
  \frac{en_0c}{j_0}=\frac{\omega_p^2}{\omega_B}\,\frac{l_0}{c}
  =\frac{l_0}{\sigma_0^{1/2}\de}\gg1
\end{equation}
is a measure of how thin the current sheets can become before approaching charge starvation. 
In particular, a current sheet supporting a jump $\delta B\sim B_0$ can collapse to the scale $\sim \sigma^{1/2} c/\omega_p$. This scale also equals the Larmor radius of the particles heated by reconnection, $r_{\rm L}\sim \sigma c/\omega_B=\sigma^{1/2}\de$. 

    \item
 The hydrodynamic velocity field of the electron-positron fluid is dominated by motions in the $x$-$y$ plane, transverse to $\vec{B}_0$. 
In particular, we observe fast motions along the current sheets, which are clearly powered by reconnection: 
plasmoids are formed by the tearing instabilities and ejected along the sheets with speeds $v_\perp^{\rm out}\sim c$. 
Electrons and positrons behave as a single fluid in these motions, described as an $\vec{E}\times \vec{B}$ drift. 
Besides the plasmoid motion along the current sheets, the hydrodynamic $\vec{E}\times \vec{B}$ motions in the $x$-$y$ plane have a component converging toward the current sheets, $\vec{v}_\perp^{\rm in}$. 
These converging flows are very fast (close to $c$) during the \textit{collapse} phases forming the flat, thin current sheets between magnetic flux ropes. 
When the sheet has formed, reconnection proceeds through the tearing instability, sustaining a significant $v_\perp^{\rm in}$.
\end{enumerate}

The current sheets are sites of strongly localized dissipation.
The degree of this localization (i.e. the spatial intermittency of dissipation) may be described by the following dimensionless parameter:
\begin{equation}\label{eq:diss}
    \mathcal{D}_J = \frac{|J|\; \vec{E}\cdot\vec{J}}{ \langle J^2 \rangle \sqrt{\langle E^2 \rangle} },
\end{equation}
where $\langle \ldots \rangle$ denotes the spatial average over the computational box. 
The map of $\mathcal{D}_J$ is shown in Figure~\ref{fig:visuals}. 
It demonstrates that the dissipation process is strongly localized in the thin current sheets. 
Taking the average of $\mathcal{D}_J$ over the simulation box, we obtain a global measure of localization,
\begin{equation}
\langle \mathcal{D}_J \rangle 
    = \frac{ \langle~ |J|\; \vec{J} \cdot \vec{E} ~\rangle} {\langle |J|\rangle\; \langle \vec{J}\cdot\vec{E} \rangle }
    = \frac{ V\, \int |J|\; \vec{J} \cdot \vec{E}\; dV} {\int |J| \,dV \, \int \vec{J}\cdot\vec{E}\; dV},
\end{equation}
where $V$ is the volume of the simulation domain. A uniform dissipation mechanism would give $\langle \mathcal{D}_J \rangle = 1$.
Instead, our fiducial 2D run with $\sigma_0 \approx 16$ gives $\langle \mathcal{D}_J \rangle \approx 3.1 \pm 0.5$ throughout an extended period of time $t>l_0/c$. 
A similar high $\langle \mathcal{D}_J \rangle$ is found in our 3D simulation and other 2D simulations with different $\sigma_0$.
The high $\langle \mathcal{D}_J \rangle$ is a clear signature of strong current-sheet dissipation.

\begin{figure}
\centering
    \includegraphics[trim={0.0cm 0.0cm 0.0cm 0.0cm}, clip=true, width=0.48\textwidth]{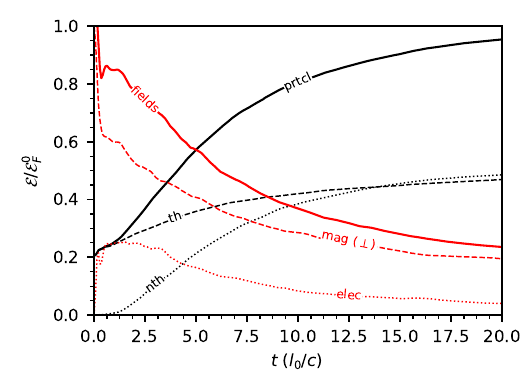}
    \caption{\label{fig:ene_nocool}
    Evolution of different energy components in the fiducial 2D simulation with  $\sigma_0\approx16$ and no radiative cooling: 
all particles (solid black), thermal particles (dashed black), nonthermal particles (dotted black), total electromagnetic field (solid red), transverse magnetic field $\vec{B}_\perp$ (dashed red), and electric field (dotted red). 
    All the components are in units of the initial total electromagnetic energy $\eneFz$.
    }
\end{figure}

\subsection{Energy Partitioning}

Turbulence transfers its energy to the plasma over the course of the simulation. 
The history of this transfer in our fiducial 2D run with $\sigma_0\approx16$ is shown in Figure~\ref{fig:ene_nocool}. 
We observed a similar history in the 3D simulation, except that the 3D turbulence is developed and the initial energy is dissipated faster by a factor of $\sim 3$.

The initial quick drop in magnetic energy at $t\lesssim l_0/c$ is mainly the result of exciting electric fields, which begin to drive MHD motions with $\vec{v}_D=c\, \vec{E}\times\vec{B}/B^2$ in response to the created magnetic stresses.
\footnote{
As one can see in Figure~\ref{fig:ene_nocool}, the injected perpendicular magnetic field energy $\eneBperp$ immediately drops by almost $40\%$. 
$25\%$ converts to the electric field energy $\eneE$, $10\%$ is given to fluctuations $\delta B_z$, and $\lesssim 5$\% goes to the kinetic energy of the excited bulk motions of the plasma.
We observed similar initial partitioning of the injected energy in simulations with  different $\sigma_0$ from $5$-$30$.
}
The ensuing gradual transfer of the turbulence energy to particles operates in two ways: 
by heating the Maxwellian pool and by accelerating nonthermal particles to high energies.

The nonthermal population grows through sudden injections from the thermal pool, as a result of energetic kicks to particles in the current sheets. 
This allows us to identify the nonthermal particles using the tracking technique described in \citet{Comisso_2018} and \cite{Nattila_2019}. 
We track the individual particle trajectories and monitor their Lorentz factors $\gamma(t)$ to detect sudden acceleration events. 
When $\dot{\gamma}=\Delta \gamma / \Delta t$ exceeds an empirical threshold of $\dot{\gamma}_{\mathrm{thr}} = 0.025 \sqrt{\sigma_0} \omp$, the particle is labeled as nonthermal (initially all particles are thermal). 
For simplicity, particles that have experienced the kick remain affiliated with the nonthermal pool until the end of the simulation, regardless of the particle history after the kick.

Energy stored in the thermal and nonthermal populations in the entire box, $\enePth(t)$ and $\enePnth(t)$, grow with time. 
Note that the \textit{thermal} part $\enePth$ effectively includes the contribution from the bulk kinetic energy; 
when needed, the latter can be separated from $\enePth$ by calculating the local bulk speed. 
The total plasma energy,
\begin{equation}
    \eneP = \enePth + \enePnth, 
\end{equation}
grows as more electromagnetic energy is dissipated.

Turbulence energy in a magnetically dominated plasma ($\sigma\gg 1$) is dominated by the electromagnetic field.
Therefore, we describe it as the total electromagnetic energy in the computational box with subtracted energy of the background field $\vec{B}_0$,
\begin{equation}
    \eneF \equiv \int \frac{E^2+B^2}{8\pi}\;dV - \frac{B_0^2}{8\pi}\,V.
\end{equation}
We observe in the simulation that $\eneF$ decays as $\propto t^{-1}$ at $t>l_0/c$, as expected for non-helical initial conditions \citep{Biskamp1999}. 
Correspondingly, $\enePth$ and $\enePnth$ grow with time.

The history of energy transfer is shown in Figure~\ref{fig:ene_nocool} for our fiducial 2D run with $\sigma_0=16$. 
At time $t=20\,l_0/c$, about $75\%$ of the initial turbulence energy $\eneFz$  has converted to plasma energy, and then $\enePth$ and $\enePnth$ remain practically unchanged (we have verified this by running the simulation up to $t \approx 100\,l_0/c$).
The thermal and nonthermal populations display two distinct distribution functions $dN_{\rm th}/d\gamma$ and $dN_{\rm nth}/d\gamma$, which grow during the main dissipation phase $t\lesssim 10\,l_0/c$ and saturate at $t\sim 20\,l_0/c$. 
Figure~\ref{fig:pspec} shows the evolution of the two distributions. 
The nonthermal component develops a broad distribution and, remarkably, the remaining thermal component is nearly Maxwellian, validating the technique for separating the two populations.

\subsection{Two-stage Nonthermal Particle Acceleration Mechanism}\label{sect:prtcl_acc}

\begin{figure}
\centering
    \includegraphics[trim={0cm 0.5cm 0cm 0.0cm}, clip=true, width=0.47\textwidth]{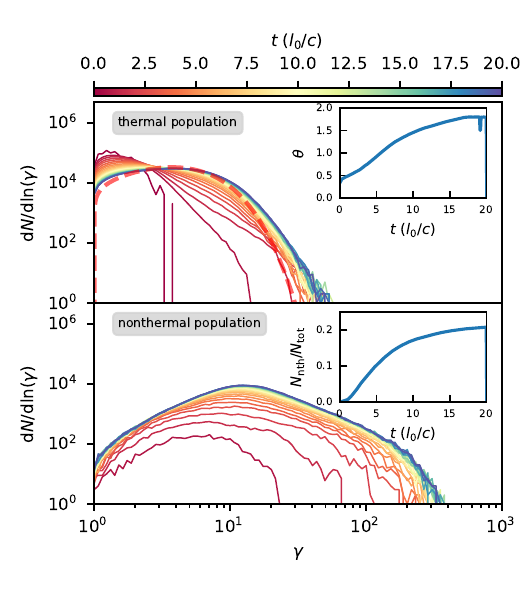}
\caption{\label{fig:pspec}
    Evolution of the particle spectra in the fiducial 2D simulation with $\sigma_0\approx16$ and no radiative cooling.
    Particles are separated into thermal (top panel) and nonthermal (bottom panel) populations, based on their acceleration histories (see text).
Thick dashed red curve shows the Maxwell-J\"uttner fit of the thermal distribution at $t \approx 20~l_0/c$.
We used similar fitting at other times to evaluate the  evolution of temperature $\theta = kT/m_e c^2$, shown in the top-right inset. 
    Bottom-right inset shows the evolution of the particle number  in the nonthermal component $N_{\rm nth}$ normalized to the total particle number in the box, $N_{\rm tot}$.
}
\end{figure}

The nonthermal particle acceleration occurs in two stages: 
the particles first receive sudden kicks and then engage in the gradual process of stochastic acceleration \citep{Comisso_2018, Comisso_2019}.
The kicks are powered by magnetic reconnection in the current sheets. 
This process \textit{injects} particles into the nonthermal population with a typical Lorentz factor
\begin{equation}
    \ginj \sim \sigma_0.
\end{equation}
The injection timescale is comparable to the time it takes to travel across a current sheet,
\begin{equation}
    \tinj \sim \sigma_0 \omega_B^{-1}.
\end{equation}

The subsequent stochastic diffusive acceleration occurs as the gyrating particles scatter off magnetic perturbations, similar to the original \citet{Fermi_1949} acceleration picture.
In this process, the particles are energized by the electric field $\vec{E}=-\vec{v}\times\vec{B}/c$ induced by the turbulent motions described in the ideal MHD approximation.
The timescale for this process depends on the exact nature of the wave-particle interaction.
For a non-resonant diffusive acceleration the timescale is comparable to the light-crossing time of the scale $l_0$ at which $\delta B/B\sim 1$ \citep[see also][]{Comisso_2019}
\begin{equation}
    \tacc \sim \frac{l_0}{c}.
\end{equation}
The acceleration process continues to be effective until the particle's Larmor radius becomes comparable to the length of the outer scale turbulent fluctuations 
\begin{equation}\label{eq:g0}
    \frac{\gamma_0 m_e c^2}{eB_0} = l_0.
\end{equation}
In our fiducial 2D simulation, $\gamma_0\approx500$.
We have verified the limit~\eqref{eq:g0} numerically:
as the simulation domain size is doubled the maximum Lorentz factor, $\mathrm{max}\{\gamma \} \sim \gamma_0$, also doubles, as expected.

A high-energy tail in the particle distribution at $\gamma \gg \sigma_0$ is expected from stochastic acceleration. 
The acceleration process can be modeled by the diffusion coefficient in the energy space $D_{\gamma}$, which may be written as 
\begin{equation}\label{eq:diff}
    D_{\gamma} \approx \zeta \frac{c}{l_0} \gamma_0^{2} \left( \frac{\gamma}{\gamma_0} \right)^{\psi}.
\end{equation}
The numerical factor $\zeta \sim 1$ controls the efficiency of stochastic acceleration.
The exponent $\psi$ is controlled by the details of the acceleration process.
The recent work by \citet{Wong_2019} reports $D_{\gamma} \propto \gamma^2$ at high $\gamma$ (i.e., $\psi = 2$).
Similarly, \citep{Comisso_2019} assume $\psi \approx 2$.  
The more accurate value of $\psi$ may be somewhat below $2$ \citep[see also the discussion in][]{Lemoine_2019, Demidem_2020}.

In agreement with previous work, the high-energy particles in Figure~\ref{fig:pspec} form a power-law distribution with spectral slope $p = d\log N/d\log \gamma \approx -2.8$.
The exact value of the slope, however, varies slightly as a function of time and magnetization (see \cite{Comisso_2019} for details).
As time progresses, more and more particles get injected into this process (bottom inset in Figure~\ref{fig:pspec}).
At the end of the simulation ($t = 20~l_0/c$) about $20\%$ of the particles reside in the nonthermal pool.

Particle energization by turbulence is an anisotropic process \citep{Zhdankin_2017a, Comisso_2019}, and we observe an anisotropic particle distribution.
Nonthermal particles with moderate $\gamma\sim\sigma_0$ are accelerated by magnetic reconnection along the guide field (which dominates over the reconnecting field component $\delta B$ on small scales). 
Particles accelerated stochastically by the \textit{motional} field $\vec{E}=-\vec{v}\times\vec{B}/c$ develop the opposite tendency: 
their angular distribution
is somewhat concentrated toward the plane perpendicular to $\vec{B} \approx \vec{B}_0$.

\section{Radiative Turbulence}\label{sect:radturb}

\subsection{Cooling Model}\label{sect:cooling_model}

The two main cooling processes in high-energy astrophysical plasmas are synchrotron emission and inverse Compton (IC) scattering of soft background photons (which can include the locally produced synchrotron photons and external radiation). 
As long as the target photons are sufficiently soft, the scattering occurs in the Thomson regime. 
Then both synchrotron and IC losses scale with the electron Lorentz factor as $\propto \gamma^2$. 
Synchrotron losses have, however, a special feature: 
they are reduced when the electrons have small pitch angles with respect to the magnetic field.

As a first step, we adopt in this paper the simplest model of radiative losses: 
each electron (or positron) with a four-velocity $u=\gamma\beta$ loses energy with rate $\dot{\cal E}=-(4/3) c \sigmat U u^2$, regardless of its velocity direction. 
This prescription accurately describes energy losses due to Thomson scattering in a cool isotropic radiation field with energy density $U$. The same prescription describes synchrotron losses in a magnetic field with energy density $U = B^2/8\pi$ if the particle distribution is isotropic and self-absorption is negligible. 
Particle acceleration by turbulence is however anisotropic \citep{Comisso_2019}, and so the cooling model will need to be refined when synchrotron losses dominate; this refinement is left for future work. 

The cooling rate for a particle with Lorentz factor $\gamma$ may be written as
\begin{equation}\label{eq:A}
    \dot{\gamma}_{\mathrm{c}} = -\mathcal{A} \frac{c}{l_0} \gamma^2\beta^2,
\end{equation}
where $\A$ is a dimensionless coefficient, which defines the cooling strength. 
This cooling rate is equivalent to a radiation drag force $\vec{F}_{\rm rad}$ directed opposite to the particle velocity $\vec{v}$, so that
\begin{equation}
    \dot{\gamma}_c m_e c^2=\vec{F}_{\mathrm{rad}}\cdot\vec{v}, \qquad
    \vec{F}_{\rm rad}=-\A\,\frac{m_ec^2}{l_0}\,\gamma^2 \vec{\beta}.
\end{equation}
In our PIC simulations, force $\vec{F}_{\rm rad}$ is added to the electromagnetic force acting on the particle, so that the net particle acceleration becomes
\begin{equation}
    \frac{\mathrm{d}\vec{u}}{\mathrm{d}t}=\frac{q_e}{m_e}\left( \vec{E} + \frac{\vec{v}}{c} \times \vec{B} \right) 
    +\frac{\vec{F}_{\mathrm{rad}}}{m_e}.
\end{equation}
The force is coupled to a relativistic Boris pusher similar to that in \citet{Tamburini_2010} and is available in the \textsc{Runko} framework as \texttt{BorisDrag} pusher.

\subsection{Weak-cooling and Strong-cooling Regimes}
\label{regimes}

To define the cooling regimes we will use the approximate description of turbulence dissipation based on the discussion in Section~\ref{sect:turb} \citep[see also][]{Sobacchi_2019}.
Comparable parts of the released energy are deposited into thermal and nonthermal plasma. 
Particle acceleration to $\ginj\sim\sigma_0$ in current sheets occurs impulsively along the magnetic field 
and stochastic acceleration to $\gamma\gg\ginj$  occurs with a diffusion coefficient $D_\gamma\sim (c/l_0)\gamma^2$, so that a particle with any $\gamma$ in the range $\ginj<\gamma<\gamma_0$ can double its energy on the timescale $\tacc\sim l_0/c$. 

The acceleration timescale should be compared with the cooling timescale $\tc(\gamma)$. 
For particles with $\gamma\gg 1$, it is given by 
\begin{equation}
     \tc(\gamma)\approx \frac{\gamma}{|\dot{\gamma}_c|} \sim \frac{l_0}{c\A\gamma}.
\end{equation}
Hereafter, we consider only systems with $\A\gg\gamma_0^{-1}=(c/\omega_Bl_0)$; 
otherwise radiative losses have a negligible effect at all relevant $\gamma<\gamma_0$. 
The condition $\A\gg\gamma_0^{-1}$ corresponds to $\tc(\gamma_0)\ll l_0/c$.  
It ensures that particles with the maximum Lorentz factor $\gamma_0$ radiate energy faster than they could gain it from the turbulence.

{\bf Weak cooling.}
We define the weak-cooling regime as 
\begin{equation}\label{eq:weak}
    \frac{1}{\gamma_0}\ll\A\ll\frac{1}{\sigma_0}.
\end{equation}
The condition $\A\ll\sigma_0^{-1}$ allows the formation of a stochastically accelerated tail in the electron distribution at $\gamma\gg\ginj\sim\sigma_0$. 
Cooling cuts off the stochastic acceleration at Lorentz factor $\gamma=\gc$ that can be estimated from $\tc(\gamma)\sim \tacc\sim l_0/c$. 
This gives
\begin{equation}
   \gc\sim  \frac{1}{\A}, \qquad  \ginj \ll\gc \ll \gamma_0.
\end{equation}

The weak-cooling condition $\A\ll\sigma_0^{-1}$ has another implication: thermal particles keep most of the energy received during the flare on the timescale $\tf\sim 10~l_0/c$.
Since the flare converts roughly half of the magnetic energy to heat, the thermal particles reach Lorentz factors $\gth\lesssim\sigma_0/2$. 
They do not efficiently cool because 
\begin{equation}
    \tc(\gth)\gg \frac{l_0}{c}.
\end{equation}

{\bf Strong cooling.}
We define the strong-cooling regime as 
\begin{equation} \label{eq:strong}
    \A\gg \frac{1}{\sigma_0}.
\end{equation}
In this case the stochastic acceleration is suppressed, however, impulsive acceleration to $\ginj\sim\sigma_0$ can still operate. 
Indeed, to shut off the impulsive acceleration, the cooling timescale $\tc(\ginj)$ would need to be shorter than the impulsive acceleration timescale $\tinj\sim\sigma_0\, c/\omega_B$, which would require $\A>\gamma_0/\sigma_0^2$. 
This is a very strong condition because astrophysical objects of interests have enormous $\gamma_0$. 
Our simulations have $\gamma_0\gg\sigma_0$, and we consider strong cooling that does not suppress the impulsive acceleration, 
\begin{equation} \label{eq:implusive_acc}
    \tc(\ginj)>\tinj, \qquad  \A<\frac{\gamma_0}{\sigma_0^2}.
\end{equation}
In our typical 2D and 3D simulations, $\gamma_0\approx 500$ and $\sigma_0\approx16$.
Therefore, the impulsive acceleration begins to be limited by radiative losses at $\A\sim 10$. 

In the strong-cooling regime, radiative losses limit the plasma temperature. 
The average thermal particle momentum during the flare, $u_{\rm th} m_ec$, is determined by the balance between cooling and heating: 
$\A\, u_{\rm th}^2 c/l_0=\dot{\gamma}_{\rm th}$, where $\dot{\gamma}_{\rm th}\sim 0.1\sigma_0 c/l_0$ is the characteristic heating rate in the flare.
This gives%
\footnote{%
Alternatively, an empirical 
formula 
from fitting our 2D simulations gives
$\dot{\gamma}_{\mathrm{th}} \sim (1/6) \sqrt{\sigma_0} c/l_0$ in a range of $8<\sigma_0<30$. 
Then $u_{\rm th}\sim 0.4\,\sigma_0^{1/4}\A^{-1/2}\ll \sigma_0$.
}
\begin{equation} \label{eq:uth}
  u_{\rm th}\sim 0.3\,\left(\frac{\sigma_0}{\A}\right)^{1/2} \ll \sigma_0.
\end{equation}

Note that $\A/\sigma_0=(2/3)(U/U_B)\tau_{\rm T}$, where $\tau_{\rm T}=\sigmat n l_0$ is the Thomson scattering optical depth of the plasma on the driving scale $l_0$. 
In radiatively efficient flares, the radiation density is comparable to $U_B$. 
Therefore, the system is expected to become optically thick if $\A\gg\sigma_0$. 
Radiative turbulence in optically thick plasma was studied by \citet{Zrake_2019} and is not considered here. 
We study cooling models with $\A<\sigma_0$. 
This implies that $u_{\rm th}$ is at least mildly relativistic. 
Note also that the relation $\sigma_0/\A\sim \tau_{\rm T}$ allows one to rewrite Equation~(\ref{eq:uth}) as $u_{\rm th}^2\tau_{\rm T}\sim 0.1$. 
Within a numerical factor, this condition is similar to the thermal balance discussed by \citet{Uzdensky_2018}.

Next, let us consider the more general model of stochastic acceleration with the diffusion coefficient $D_\gamma$ given in Equation~\eqref{eq:diff}.
The approximation $D_\gamma\sim \gamma^2 c/l_0$ corresponds to $\psi=2$ and $\zeta\sim 1$. 
The actual value of $\psi$ may be slightly below 2, and this has an important consequence, as seen from the following.
The average rate of stochastic acceleration at a given $\gamma$ is 
\begin{equation}
    \dot{ \langle \gamma \rangle } = \frac{\mathrm{d} D_{\gamma}}{\mathrm{d} \gamma} = \zeta \psi \omega_B \left( \frac{\gamma}{\gamma_0} \right)^{\psi-1},
\end{equation}
and the corresponding acceleration timescale is
\begin{equation}
\label{eq:tacc}
    \tacc = \frac{\gamma}{ \dot{\langle \gamma \rangle} } = \frac{l_0/c}{\zeta \psi} \left( \frac{\gamma}{\gamma_0} \right)^{2-\psi}.
\end{equation}
Radiative cooling stops acceleration where $\tc(\gamma)$ becomes equal to $t_{\mathrm{acc}}(\gamma)$. This occurs when the particle reaches
\begin{equation}
\label{eq:gc}
   \gc \sim \gamma_0 \left( \frac{\zeta\, \psi}{\A\,\gamma_0} \right)^{\frac{1}{3-\psi}}.
\end{equation}
Comparing $\gc$ with $\ginj\sim\sigma_0$, one finds
\begin{equation} \label{eq:dissmeasure1}
    \frac{\gc}{\ginj}\sim \left( \frac{\zeta\, \psi}{\A\,\sigma_0} \right)^{\frac{1}{3-\psi}} \left(\frac{\gamma_0}{\sigma_0}\right)^{\frac{2-\psi}{3-\psi}}.
\end{equation}
When $\psi<2$, $\gc$ can significantly exceed $\ginj\sim\sigma_0$, even when $\A>\sigma_0^{-1}$, because $\gamma_0$ is typically very large. 
This means that stochastic acceleration beyond $\ginj$ can be efficient even in the strong-cooling regime (when the plasma temperature is limited by radiative losses). 
In particular, $\gamma_0$ in bright accreting black holes and their jets exceeds $10^{10}$ \citep{Beloborodov_2017}, while typical $\sigma_0$ may be, e.g. $\sim 10$. 
Then $\psi=1.7$ gives $(\gamma_0/\sigma_0)^{(2-\psi)/(3-\psi)}\sim 10^2$. 
Thus, the modest reduction of $\psi$ from 2 to 1.7 enables stochastic acceleration well above $\ginj$.

Finally, we note that strong radiative losses could in principle suppress the turbulent bulk motions. 
The drag force damping a given turbulent eddy is given by 
\begin{equation}
  \vec{F}_{\rm drag}=-\A\,\frac{m_e c^2}{l_0}\overline{u^2}\,\overline{\vec{\beta}},
\end{equation}
where the bar signifies averaging over the particle motions in the eddy. 
The effective inertial mass per particle is $\sim\sigma_0m_e\gg m_e$, and the timescale for damping the turbulent bulk momentum $\overline{\vec{u}}\sim\sigma_0 m_ec\overline{\vec{\beta}}$ is 
\begin{equation}
   t_{\rm drag}=\frac{|\overline{\vec{u}}|}{F_{\rm drag}}=\frac{\sigma_0\, l_0}{\A\, c\, \overline{u^2}}
   \sim 10\,\frac{u_{\rm th}^2}{\overline{u^2}}\,\frac{l_0}{c}.
\end{equation}
Assuming that the impulsive acceleration of a fraction of particles to $\ginj\sim\sigma_0$  (followed by fast cooling of the particles) does not greatly increase $\overline{u^2}$, so that $\overline{u^2}\sim u_{\rm th}^2$, one concludes that $t_{\rm drag}$ exceeds the cascade timescale $\sim l_0/c$. 
This allows the cascade to develop, overcoming the drag force.

\subsection{Quasi-steady One-zone Acceleration Model}

Magnetic reconnection injects particles into the stochastic acceleration process with the Lorentz factor $\gamma_{\mathrm{inj}} \sim \sigma_0$ \citep{Comisso_2018}. 
Let us consider the situation where $\gc \gg \ginj$ and focus on particles with Lorentz factors $\gamma>\ginj$. 
Let us assume here that $D_\gamma$ is uniform in the turbulent plasma; 
we will call it the \textit{one-zone} model.
The particle distribution function $f(\gamma, t)$ then satisfies 
\begin{equation}
    \partial_t f + \partial_\gamma(-D_\gamma \partial_\gamma f + \dot{\gamma}_{\mathrm{c}} f)
    = \dot{n}\, \delta(\gamma - \gamma_{\mathrm{inj}}),
\end{equation}
where $\dot{n}$ is the injection rate.
On timescales longer than the average time of acceleration to $\gamma_{\mathrm{cool}}$ one can expect $f(\gamma)$ to evolve in a quasi-steady regime, when $\partial_t f$ is small compared with the other terms on the left-hand side.
Then the shape of $f(\gamma)$ at $\gamma > \gamma_{\mathrm{inj}}$ approximately satisfies 
\begin{equation}
\label{eq:flux}
    {\cal F}=-D_\gamma \frac{\ud f}{\ud \gamma} + \dot{\gamma}_{\mathrm{c}} f = 0.
\end{equation}
It states that the particle flux in the energy space  ${\cal F}$ vanishes in the steady state (a non-zero ${\cal F}$ is forbidden by the \textit{ceiling} of radiative losses at $\gamma\gg\gc$). 
The solution of Equation~\eqref{eq:flux} is 
\begin{equation}\label{eq:cutoff}
    f(\gamma) = K(t) \exp \left[ -\left( \frac{\gamma}{\gamma_{\mathrm{cool}}} \right)^{3-\psi} \right],
\end{equation}
where
\begin{equation}
    \gamma_{\mathrm{cool}}^{3-\psi} = \frac{(3-\psi)\zeta}{\mathcal{A}} \gamma_0^{2-\psi}.
\end{equation}
This definition of $\gc$ is equivalent to Equation~(\ref{eq:gc}) within a numerical factor $\sim 1$. The normalization constant $K(t) \sim \dot{n}t$ is growing on the timescale longer than $t_{\mathrm{acc}}(\gamma_{\mathrm{cool}})$.
Its more accurate value is given by
\begin{equation}
    K(t) \approx 
    \frac{ \int_0^t \dot{n}(t') \ud t' }{ \int_{\gamma_{\mathrm{inj}}}^{\infty} \exp \left[ -\left(\frac{\gamma}{\gamma_{\mathrm{cool}}} \right)^{3-\psi} \right] \ud\gamma }.
\end{equation}

Radiative cooling offers a useful test for the one-zone description of stochastic particle acceleration with a given $D_\gamma$. 
Then, cooling generally results in a quasi-steady solution for $f(\gamma)$ (illustrated by Equation~(\ref{eq:cutoff})) with an exponential cutoff at $\gc$ and a hard slope below the cutoff, as particles injected at $\gamma\sim \ginj$ diffuse toward $\gc$ and accumulate there. 
Our simulations (presented below in Sect.~\ref{sect:results}) show that the actual distribution function in the computational domain is far different from this one-zone solution. 
This difference is caused by the strong spatial intermittency of turbulence, which results in large variations of $D_\gamma$ across the domain. 
Note also that for any given $\gamma$, $D_\gamma(\gamma)$ can be defined only in sufficiently large regions exceeding the Larmor radius $r_{\rm L}(\gamma)=l_0\gamma/\gamma_0$.

\begin{figure*}
\centering
    \includegraphics[width=0.47\textwidth]{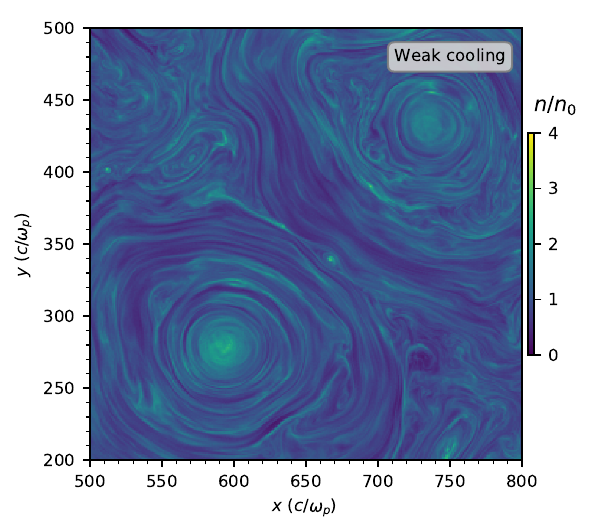}
    \includegraphics[width=0.47\textwidth]{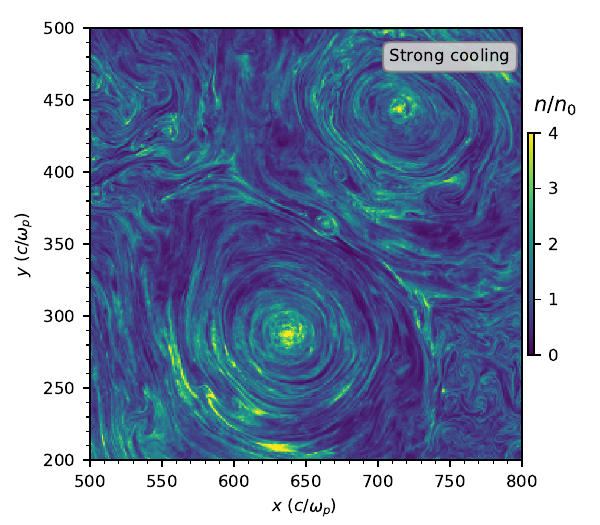}
    \caption{\label{fig:visuals_rad}
Zoom-in on the plasma density $n$ (shown in units of the initial $n_0$) in a subregion of the computational box.
The two panels show the snapshots of two simulations at time $t \approx 10 \,l_0/c$. The simulations have the same parameters (our fiducial 2D model $\sigma_0 \approx 16$) except the different cooling rates:
$\mathcal{A} = 0.001$ (left) and $\mathcal{A} = 2.6$ (right). One can see that stronger cooling results in a higher density contrast.
}
\end{figure*}

\section{Simulations of Radiative Turbulence}\label{sect:results}

We have repeated our fiducial simulation described in the previous sections, but now with radiative cooling, $\A\neq 0$. 
We used five long 2D simulations (with 
$\A \approx$
$\Ten{1.0}{-3}$, 
$\Ten{1.6}{-3}$, 
$\Ten{2.8}{-3}$, 
$\Ten{6.4}{-3}$, and
$\Ten{2.6}{-2}$) 
to explore the weak cooling regime $\gamma_0^{-1} \lesssim \A < \sigma_0^{-1}\approx 0.06$. 
In this regime, the radiative losses restrict, but do not suppress completely, stochastic particle acceleration. 
The strong cooling regime, $\A >\sigma_0^{-1}$, is covered with three 2D runs (with $\A \approx 0.10$, $0.41$ and $2.6$).
In addition, we performed shorter simulations with different magnetizations $\sigma_0$, and with $\A$ as low as $10^{-4}$ and as high as $20$.

Full 3D models are expensive, and we have done only two 3D simulations that include radiative cooling, with $\A \approx 0.003$ and $0.1$, probing the weak and strong cooling regimes, respectively; 
both setups have $\sigma_0\approx 16$.

We have found from these simulations that radiative losses weakly affect the turbulence magnetic spectrum $d{\cal E}_B/dk$; 
it is practically the same as without losses (Figure~\ref{fig:spec}). 
Nevertheless, strong losses impact the perturbations of plasma density, as clearly seen in Figure~\ref{fig:visuals_rad}. 
While the radiative drag force $\vec{F}_{\rm rad}$ is unable to damp the turbulent motions of the magnetic field lines it significantly affects the plasma motion {\it along} the field lines, which leads to accumulation of the high-density stripes. 
In the strong cooling regime, we also observed that the thermal plasma comes to a quasi-steady temperature, at which cooling balances heating. 
Approaching this balance took $\sim 10\,l_0/c$ in  2D and $\sim 3\,l_0/c$ in  3D simulations. 
The radiative turbulent flares have less relativistic temperatures (and hence higher effective magnetizations $\sigma$) compared with turbulence without cooling.

\subsection{Radiation Effect on Particle Acceleration}

\begin{figure}
\centering
    \includegraphics[trim={0cm 0.6cm 0cm 0.2cm}, clip=true, width=0.43\textwidth]{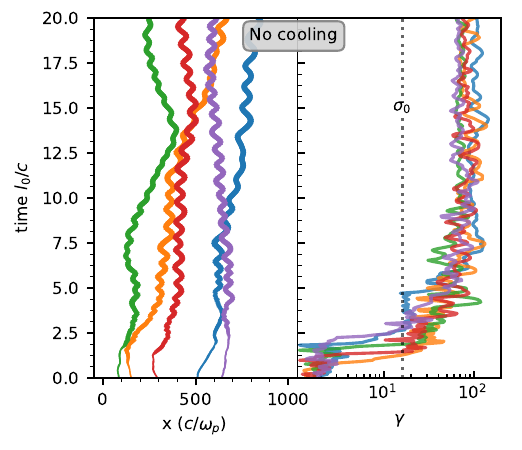}
    \includegraphics[trim={0cm 0.6cm 0cm 0.2cm}, clip=true, width=0.43\textwidth]{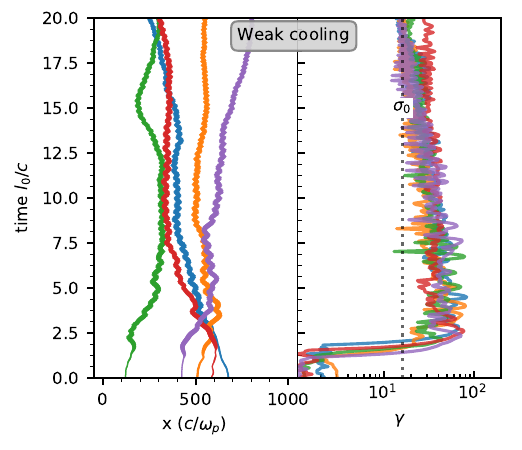}
    \includegraphics[trim={0cm 0.2cm 0cm 0.2cm}, clip=true, width=0.43\textwidth]{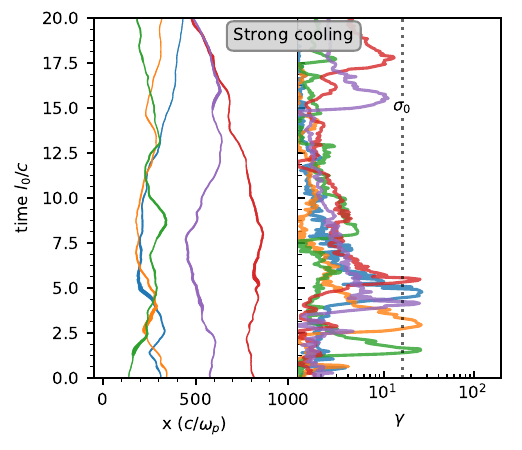}
\caption{\label{fig:prtcl_energization_history}
    Energization histories of five representative high-energy particles tracked in the 2D simulations with $\sigma_0 \approx 16$, with $\mathcal{A}=0$ (top), $\mathcal{A} \approx 0.003$ (middle), and $\mathcal{A} \approx 0.1$ (bottom).
    Left panels show the particle $x$ coordinate, and right panels show the particle Lorentz factor $\gamma$ (line width in left panels is proportional to $\gamma$).
    Vertical dotted line shows $\gamma\sim\sigma_0$ -- the expected gain from an injection event in a reconnecting current sheet.
    After an initial kick at a reconnection site the particles enter the stochastic acceleration process where they gyrate and scatter between the magnetic islands.
    In the weak-cooling regime the particles reach the radiative cooling limit $\gc\gg \sigma_0$, which slowly decreases as the turbulence decays.
    In the strong-cooling regime, the particles remain cool except for short acceleration events, which happen when the particles enter a reconnecting current sheet.
}
\end{figure}

\begin{figure*}
\centering
    \includegraphics[trim={0cm 0.5cm 0cm 0.0cm}, clip=true, width=0.46\textwidth]{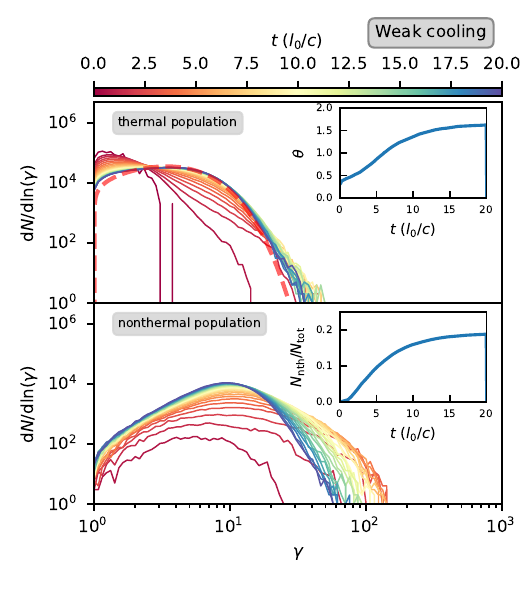}
    \includegraphics[trim={0cm 0.5cm 0cm 0.0cm}, clip=true, width=0.46\textwidth]{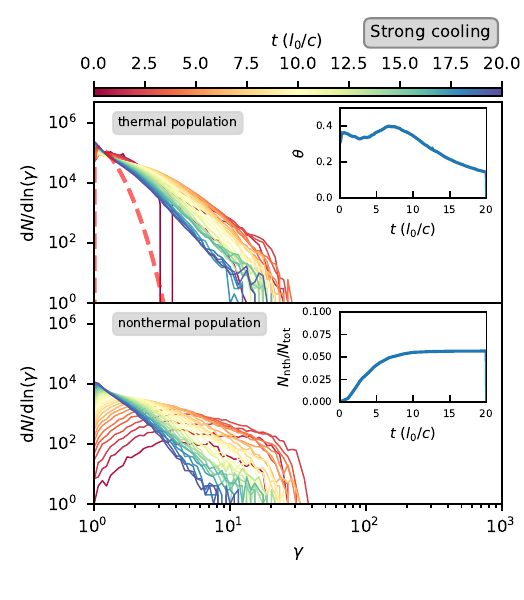}
\caption{\label{fig:pspec_weak}
    Evolution of the particle spectra in the fiducial 2D simulation  with $\sigma_0 \approx 16$ with weak cooling ($\A \approx 0.003$, left) and strong cooling ($\A \approx 0.1$, right).
    Symbols and colors are the same as in Figure~\ref{fig:pspec}.
}
\end{figure*}

Figure~\ref{fig:prtcl_energization_history} shows the acceleration histories of five representative high-energy particles picked in the computational box, for three simulations with no cooling ($\A = 0$), weak cooling ($\A \approx 0.003$), and strong cooling ($\A \approx 0.1$), all for the fiducial 2D turbulent flare with $\sigma_0 \approx 16$.

As expected, the weak cooling regime preserves the two-stage acceleration process: 
impulsive  acceleration to $\gamma \sim \sigma_0$ at reconnection sites is followed by slow stochastic acceleration to higher energies.
Then the particles reach a balance between acceleration and cooling with Lorentz factors of $\gamma \approx \gamma_{\mathrm{cool}}$.
The equilibrium $\gamma_{\mathrm{cool}}$ gradually decreases as the turbulence dissipates. 
By contrast, in the strong-cooling regime, only the impulsive acceleration remains active.
The accelerated particles quickly cool down back to $\sim\gth$ when they exit the reconnection site with
the accelerating $\vec{E}_\parallel$. 
Some particles experience multiple kicks by $\vec{E}_\parallel$ in regions with active current-sheet formation; however, the slow stochastic acceleration is completely suppressed.

The evolution of the particle spectrum during the turbulent flare (measured in the entire computational box) is shown in Figure~\ref{fig:pspec_weak}. 
We have used the same particle tracking technique as in Figure~\ref{fig:pspec} to disentangle the particle distribution into the thermal and nonthermal components.

In the weak-cooling regime, we observe the extension of the nonthermal spectrum to $\gc$;
radiative losses exponentially suppress the particle population at $\gamma>\gc$. 
The value of $\gc$ can be measured by fitting the particle distribution with the exponential cutoff, and we find $\gamma_{\mathrm{cool}} \approx 100$ during the main phase of the turbulent flare, when stochastic acceleration is strongest. 
The nonthermal particle fraction ($\approx 20\%$) and the slope of the nonthermal particle spectrum ($p \approx -2.5$) are then similar to those in the simulation without cooling (compare the left panel in Figure~\ref{fig:pspec_weak} with Figure~\ref{fig:pspec}).
At later times, $\gc$ decreases, as the turbulence decays. 
At the end of the simulation ($t = 20\,l_0/c$) $\gc\sim 10$ becomes comparable to both $\gth$ and $\ginj\sim \sigma_0$. 
Almost all the remaining energy is then stored in the hot thermal plasma.

In the strong-cooling regime, the particle spectrum shows a suppression of the nonthermal population.
Particles still experience impulsive acceleration events at reconnection sites; 
however, their fast cooling keeps the nonthermal particle fraction $N_{\mathrm{nth}}/N_{\mathrm{tot}}$ low, below 6\%.
\footnote{
    Our decomposition of particles into thermal and nonthermal populations becomes less accurate in the strong-cooling regime because of finite time sampling. 
    Strong energy losses make it harder for the tracked particles to exceed the injection threshold $\dot{\gamma}_{\mathrm{thr}}$ for long periods of time.
    This leads to missing of some injected particles that should in reality be marked as belonging to the nonthermal population. 
}
The plasma temperature remains roughly constant until $t \approx 8 \,l_0/c$, in the heating=cooling balance. 
Then the heating rate decays, and the plasma temperature decreases.

\begin{figure}
\centering
    \includegraphics[width=0.48\textwidth]{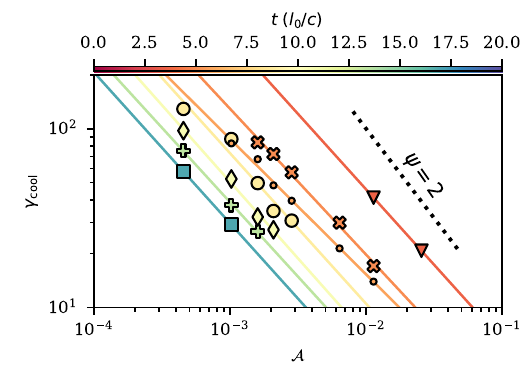}
\caption{\label{fig:gcool_s10}
Dependence of $\gc$ on the cooling parameter $\A$. 
We ran simulations of the fiducial 2D model ($\sigma_0\approx 16$) with different values of $\A$, and analyzed the particle energy distribution in various snapshots of these simulations to measure $\gc(\A,t)$.
Different colors/symbols in the figure correspond to  different times $t$, as indicated by the color bar above the figure. 
Colored lines show the fits of $\gc\propto\A^w$ at seven different times. The fit slope $w$ determines $\psi=3+w^{-1}$. 
The black dotted line indicates the slope $\psi = 2$.
}
\end{figure}

\subsection{Measuring Stochastic Acceleration Rate Using $\gc$}\label{sect:psi}

Since acceleration is balanced by the (known) radiative losses at $\gamma=\gc$, the measurement of $\gc$ gives the acceleration rate at $\gamma=\gc$. 
Note that the same turbulent flare can have different $\gc$, depending on $\A$. By varying $\A$ we can vary $\gc$ and thus find how the particle acceleration rate scales with $\gamma$. 
In the standard picture of stochastic acceleration, particles gain energy as a result of diffusion in the energy space. 
Using the measured $\gc(\A)$, one can evaluate the exponent $\psi$ of the diffusion coefficient  $D_\gamma\propto\gamma^\psi$. 
The relation  $\gc\propto \A^{-1/(3-\psi)}$ (Equation~\ref{eq:dissmeasure1}) offers a convenient way to find $\psi$.
We performed the measurements of $\gc$ for a range of $\A$ that give $\gc>\ginj\sim\sigma_0$ (the weak-cooling regime), in which stochastic acceleration is not suppressed.
The measurement is made by fitting the nonthermal particle spectrum by a power law with an exponential cutoff.
We use our fiducial 2D model of a turbulent flare with $\sigma_0\approx 16$. We run the model with different $\A$ and take snapshots of the runs at equal times (so that we compare stochastic acceleration at the same turbulence level). 
The results are shown in Figure~\ref{fig:gcool_s10}.

A good time interval for measuring $\gc(\A)$ is $5 < c t/l_0 < 10$.
Earlier times $t$ are not suitable in the simulations with low $\A$ (because it takes time for the particles to establish the extended high-energy tail with at a high $\gc$), and later times are not suitable in the simulations with high $\A$ (because $\gc$ drops to $\ginj$, as the turbulence decays). 
This measurement gives $\psi = 1.68 \pm 0.13$.

This procedure of measuring $\psi$ assumes that the stochastic acceleration can be described by a universal scaling of the diffusion coefficient $D_\gamma\propto\gamma^\psi$ (with a prefactor determined by the turbulence level). 
As we show below, this standard picture is in fact deficient because the stochastic acceleration is strongly intermittent in space and time.

\begin{figure*}
\centering
    \includegraphics[trim={0cm 0.1cm 0cm 0.2cm}, clip=true, width=0.45\textwidth]{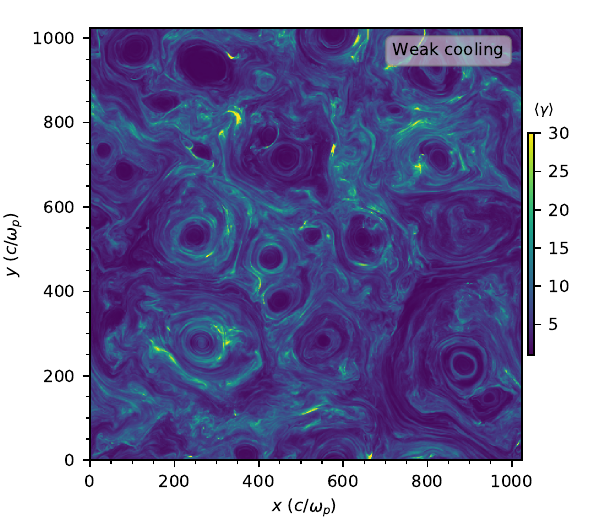}
    \includegraphics[trim={0cm 0.1cm 0cm 0.2cm}, clip=true, width=0.45\textwidth]{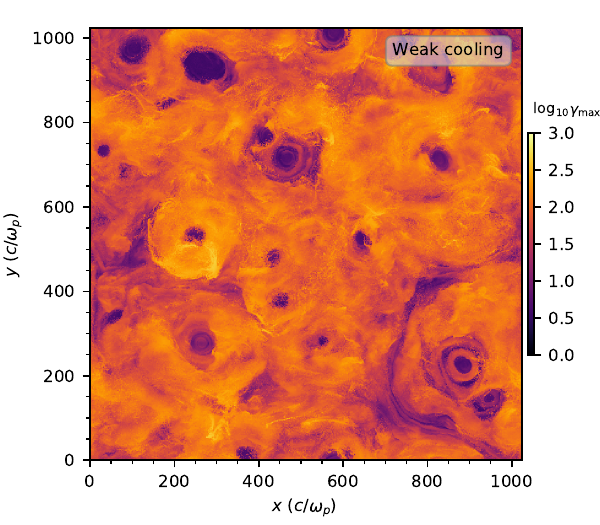}
    \includegraphics[trim={0cm 0.1cm 0cm 0.2cm}, clip=true, width=0.45\textwidth]{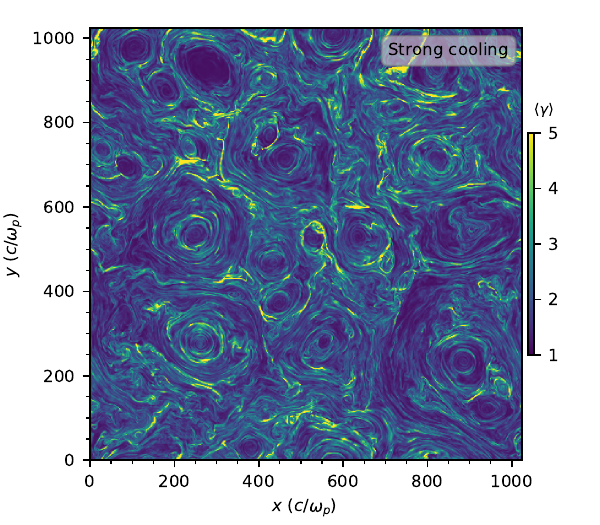}
    \includegraphics[trim={0cm 0.1cm 0cm 0.2cm}, clip=true, width=0.45\textwidth]{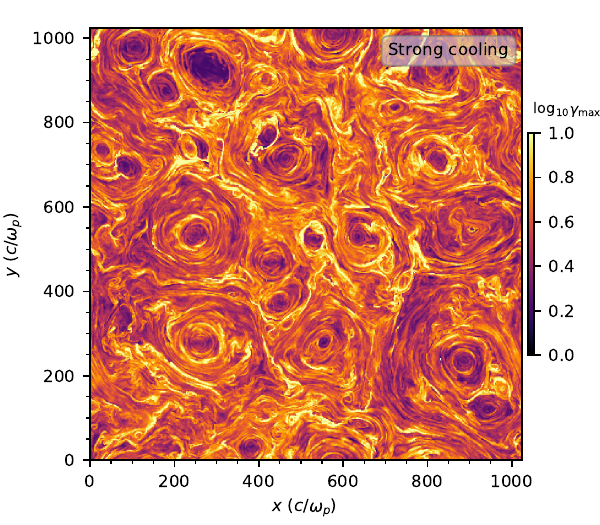}
\caption{\label{fig:interm}
    Visualization of the intermittency of the particle energization in the 2D simulations with a magnetization of $\sigma_0 \approx 16$ in the weak-cooling (top row; $\mathcal{A} \approx 0.003$) and strong-cooling regimes (bottom row; $\mathcal{A} \approx 0.1$) at $t \approx 10~l_0/c$.
    Left panels show the mean Lorentz factor $\langle \gamma \rangle_\mathrm{c}$, 
    and right panels the logarithm of the maximum Lorentz factor, $\log_{10} (\mathrm{max}\{ \gamma \})$.
    These quantities are computed using a set of particles located inside $2 \times 2 \,\de$ resolution pixels.
    Note the different color scales.
    In the weak-cooling regime energetic particles are preferentially located in hot streams in between the magnetic islands.
    In the strong-cooling regime the production of the nonthermal population is suppressed and thin current sheets are the main heating and particle acceleration sites.
}
\end{figure*}

\begin{figure*}
\centering
    \includegraphics[trim={0cm 0.1cm 0cm 0.2cm}, clip=true, width=0.45\textwidth]{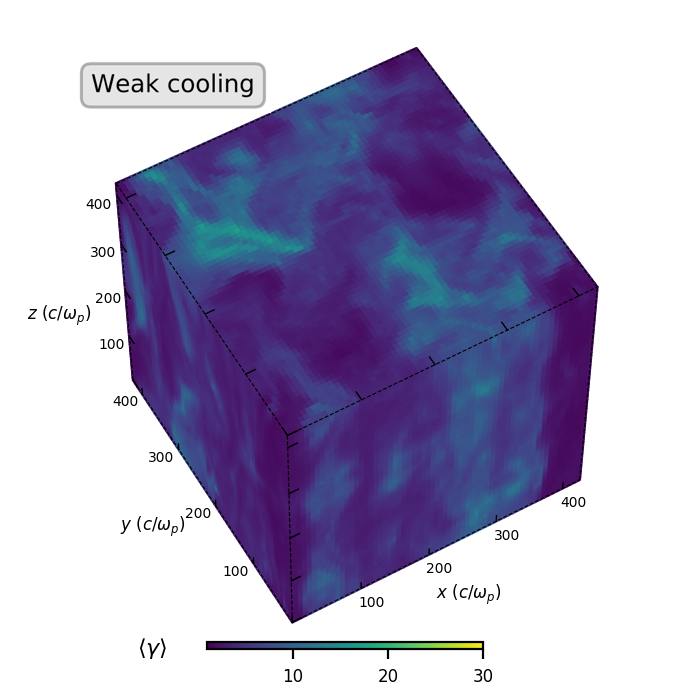}
    \includegraphics[trim={0cm 0.1cm 0cm 0.2cm}, clip=true, width=0.45\textwidth]{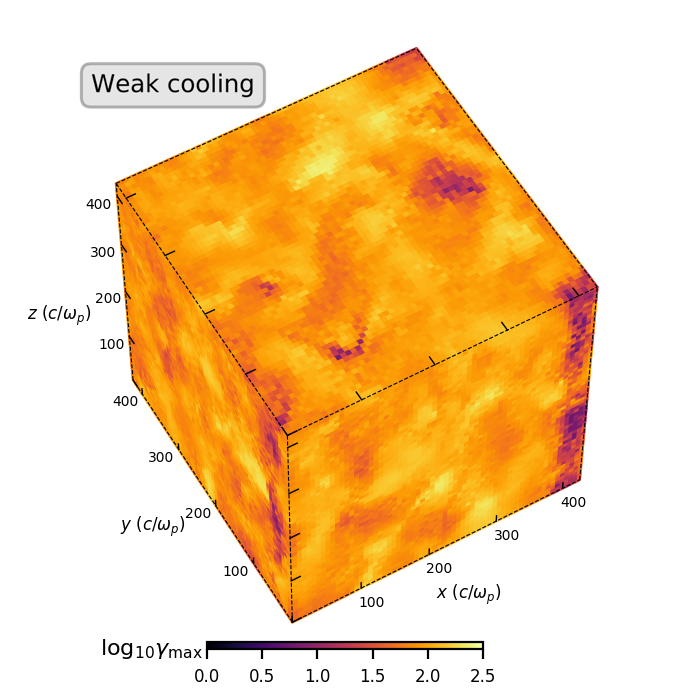}
    \includegraphics[trim={0cm 0.1cm 0cm 0.2cm}, clip=true, width=0.45\textwidth]{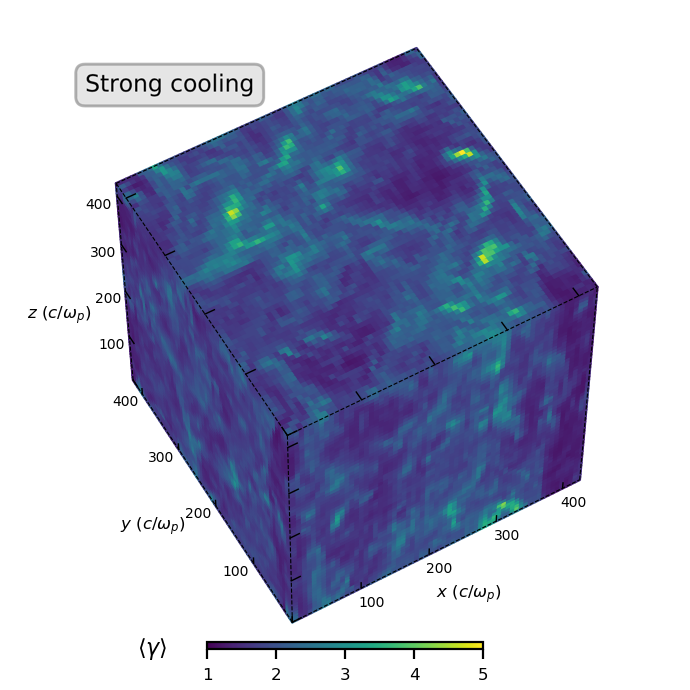}
    \includegraphics[trim={0cm 0.1cm 0cm 0.2cm}, clip=true, width=0.45\textwidth]{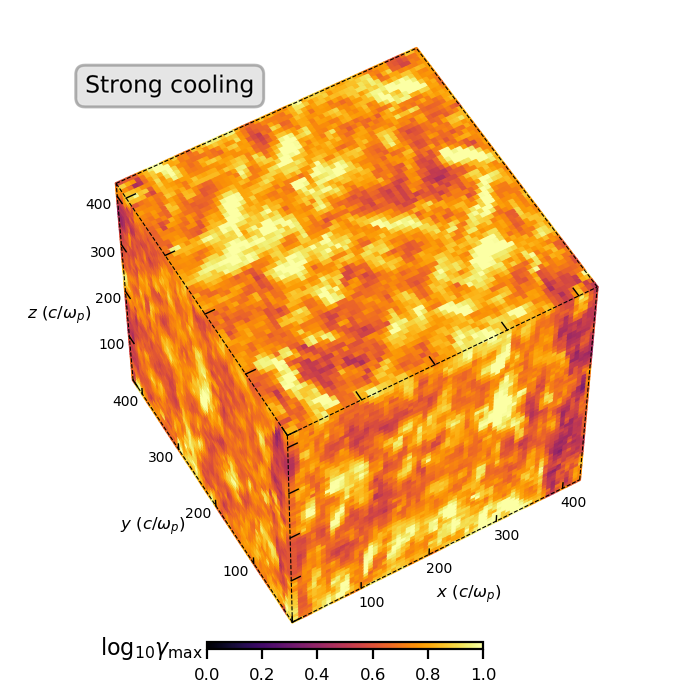}
\caption{\label{fig:interm3d}
    Visualization of the intermittency of the particle energization in the 3D simulations with a magnetization of $\sigma_0 \approx 16$ in the weak-cooling (top row; $\mathcal{A} \approx 0.003$) and strong-cooling (bottom row; $\mathcal{A} \approx 0.1$) regimes at $t \approx 5~l_0/c$.
    Both quantities are computed using a set of particles located inside $6 \times 6 \times 6 \,\de$ resolution voxels.
    Symbols are the same as in Figure~\ref{fig:interm}.
    In addition to perpendicular variations in the $xy$-plane, similar to the 2D simulation results the particle distributions form column-like structures along the guide field ($\hat{\vec{z}}$-axis) direction.
}
\end{figure*}

\begin{figure}
\centering
    \includegraphics[trim={-0.5cm 0.0cm 0cm 0.0cm}, clip=true, width=0.49\textwidth]{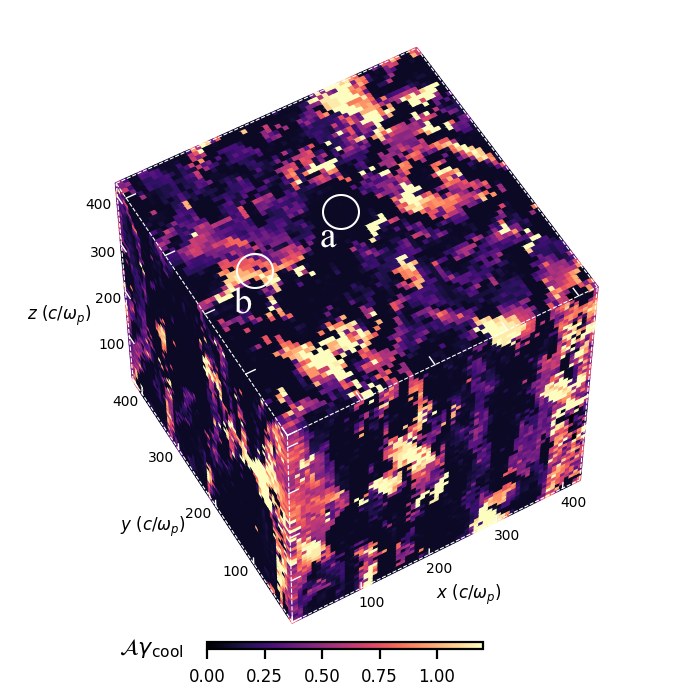}
    \includegraphics[trim={0.0cm 0.0cm 0cm 0.0cm}, clip=true, width=0.41\textwidth]{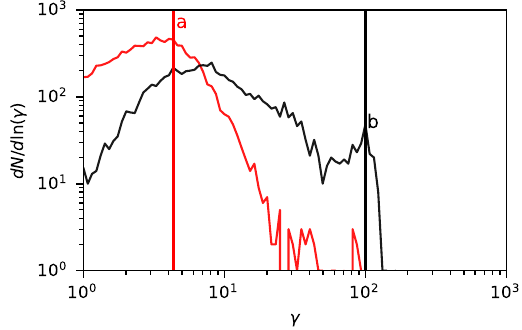}
\caption{\label{fig:acc_interm}
Spatial intermittency of stochastic particle acceleration observed in the 3D simulation with $\sigma_0 \approx 16$ and $\A \approx 0.003$.
The snapshot is taken at time $t \approx 5~l_0/c$.
Top panel shows the map of $\gc$ at which radiative losses balance the local acceleration. 
Its value is measured in each sub-domain of size $12\times12\times12 \,(\de)^3$.
Bottom panel shows the particle distributions at two locations (indicated by white circles ``a'' and ``b'' in the upper panel). 
The two vertical lines show the position of $\gc$ for the two distributions.
}
\end{figure}

\subsection{Intermittency of Particle Acceleration}\label{sect:intermittency}

The spatial intermittency of dissipation in the simulated turbulent flares was already mentioned in Section~3.1: 
the  magnetic energy release is strongly enhanced in the current sheets. 
We also observe that the stochastic acceleration of particles is highly inhomogeneous in the computational box. 
Simulations with strong radiative losses are helpful for the analysis of intermittency, as strong emission highlights the localized regions of strong heating, especially if cooling is faster than particle diffusion out of the heating region.

To study the spatial intermittency, we divided the computational box into small domains, $2\times2\,(\de)^2$ in 2D and $6\times6\times6 \,(\de)^3$ in 3D, and analyzed the local particle populations in each domain. 
A measure of heating in each domain is the mean particle Lorentz factor $\langle \gamma \rangle$. 
The simplest measure of nonthermal particle acceleration is the maximum Lorentz factor $\gamma_{\max}$.

Figure~\ref{fig:interm} shows the snapshots of $\langle \gamma \rangle$ and $\gamma_{\max}$ from two 2D simulations, with $\A \approx 0.003$ and $\A \approx 0.1$.
A similar analysis of the 3D model is shown in Figure~\ref{fig:interm3d}.
We observe that the reconnecting current sheets and their immediate vicinity are hot. 
The hot regions are particularly narrow in the simulation with $\A \approx 0.1$ --- the particles quickly cool when exiting the heating regions, and so a high $\langle \gamma \rangle$ reflects a high heating rate.
In the 2D simulations, the magnetic islands are seen to have a distinct cold inner core, as the islands are capable of insulating themselves from the hot surrounding particles.
This effect is slightly alleviated in full 3D simulations where the outer edges of the flux ropes appear less pronounced \citep[see also][]{Comisso_2019}, and the flux ropes are bent and reconnect on a scale $\sim L$ along the guide field.

Turbulence with weak cooling ($\A \approx 0.003$) enables stochastic particle acceleration, and $\gamma_{\max}$ reaches high values.
The map of $\gamma_{\max}$ in this regime is quite diffuse, as expected --- the stochastic acceleration to high $\gamma$ takes a significant time and is accompanied by significant spatial diffusion.

As explained in Section~5.2, simulations with radiative losses provide another measure of stochastic acceleration: 
$\gc$, which can be found from the observed particle spectrum.
The measurement of $\gc$ can be done locally, in each small sub-domain of the computational box. 
The results are shown in Figure~\ref{fig:acc_interm} for the 3D simulation with $\A \approx 0.003$. 
In a simple model of stochastic particle acceleration with a constant diffusion coefficient $D_\gamma$ one would expect a uniform $\gc\approx \A^{-1}$ throughout the computational box. 
By contrast, we observe a highly inhomogeneous map of $\A\gc$, with large \textit{voids} where  $\A\gc\ll 1$.
This map demonstrates that $D_\gamma$ strongly varies on scales $\gtrsim 20 \,\de$.

We conclude that although the low-energy part of the total volume-integrated particle distribution appears to be well described by a single-temperature Maxwell-J\"{u}ttner distribution (see Figures~\ref{fig:pspec} and \ref{fig:pspec_weak}), the particle
heating and acceleration are in fact strongly intermittent.
Both are suppressed away from current sheets, the sites where the dissipation peaks.

\subsection{Energy Partition} 

\begin{figure}
\centering
    \includegraphics[trim={0.3cm 0.4cm 0.0cm 0.3cm}, clip=true, width=0.46\textwidth]{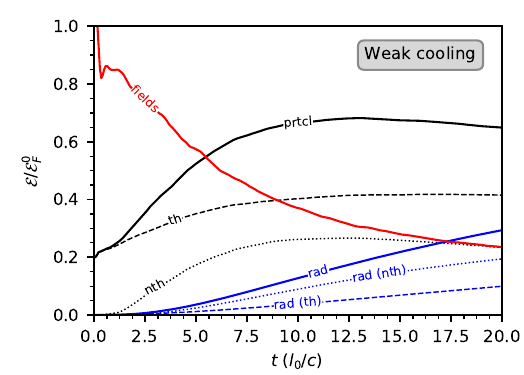}
    \includegraphics[trim={0.3cm 0.0cm 0.0cm 0.3cm}, clip=true, width=0.46\textwidth]{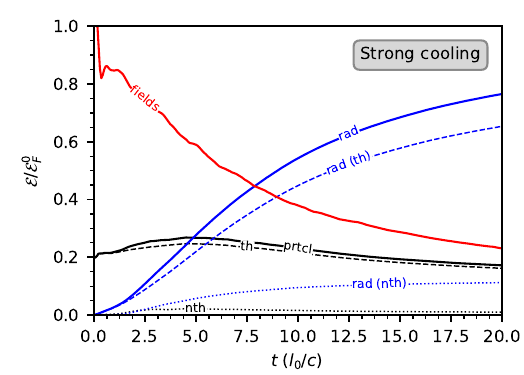}
    \caption{\label{fig:ene_cooling}
Evolution of different energy components in the fiducial 2D simulation with $\sigma_0\approx16$, with weak cooling (top panel; 
$\mathcal{A} \approx 0.003$) and strong cooling (bottom panel; 
$\mathcal{A} \approx 0.1$). 
All the components are shown in units of the initial total electromagnetic energy $\eneFz$ (cf. Figure~\ref{fig:ene_nocool}). 
The figure shows the electromagnetic energy (solid red) and the total energy of particles (solid black), which is separated into contributions from thermal particles (dashed black) and nonthermal particles (dotted black). 
The blue curves show the energy carried away by radiation.
The total radiation energy (solid blue) is the sum of the energies emitted by thermal particles (dashed blue) and nonthermal particles (dotted blue).
}
\end{figure}

\begin{figure}
\centering
    \includegraphics[trim={0.0cm 0.0cm 0.0cm 0.0cm}, clip=true, width=0.46\textwidth]{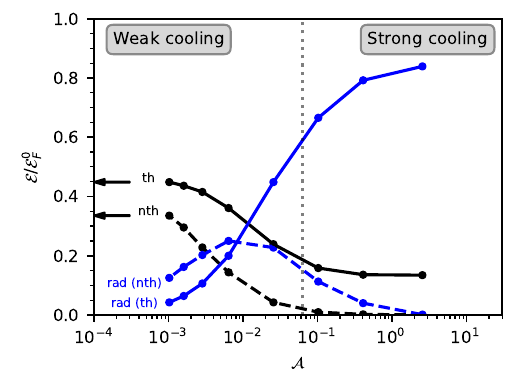}
    \caption{\label{fig:energy_budget}
Energy carried by the plasma (black) and radiation (blue) at the end of the fiducial 2D simulation with $\sigma_0=16$, at $t=20 l_0/c$. The simulation was run 8 times with 8 different cooling levels $\A>0$, and the final energy partitioning is presented as a function of $\A$.
The figure shows separately the energies carried by thermal (solid black) and nonthermal (dashed black) particles. 
The radiation energy is also separated into two parts, emitted by thermal (solid blue) and nonthermal (dashed blue) particles.
All components are normalized to the initially injected turbulence energy $\eneFz$.
Measurements from the simulation with no cooling ($\mathcal{A} = 0$) are shown by the horizontal arrows.
}
\end{figure}

Section~3.2 described the evolution of different energy components in our fiducial 2D simulation without radiative losses. 
Figure~\ref{fig:ene_cooling} shows how this evolution is changed by weak ($\A \approx 0.003$) and strong ($\A \approx 0.1$) losses. 
We also performed 3D simulations with $\A \approx 0.003$ and $0.1$, with similar results.

The electromagnetic field components evolve almost identically to the non-cooling case and show no dependency on $\A$.
At the end of the 2D simulations (at $t = 20~l_0/c$) the energy of the transverse magnetic field $\eneBperp$ has decayed to $\sim 20\%$ of the initially injected magnetic energy, and the electric field energy carries $\sim 5\%$. 
Energy in the parallel (guide) magnetic field stays nearly constant.
These numbers remain practically unchanged at $t>20l_0/c$, which was confirmed by runs up to $t\sim 100l_0/c$.

One can see from Figure~\ref{fig:ene_cooling} that in the weak-cooling regime most of the energy lost by the electromagnetic field is retained by the plasma by the end of simulation (time $t=20l_0/c$), and the radiation has been produced mostly by the nonthermal particles. 
By contrast, in the strong-cooling regime, most of the turbulence energy is lost to radiation, and the emission is dominated by thermal particles. 
This behavior is further demonstrated in Figure~\ref{fig:energy_budget}, which presents the final partitioning of the released energy between the plasma and radiation for the fiducial 2D model with 9 different cooling levels $\A$, varying from $\A=0$ to $\A\approx 3$. 
We find that $\A\gtrsim 0.02$ is sufficient to radiate away most of the turbulence energy before the end of the simulation. 
The emission becomes dominated by the thermal particles at $\A\gtrsim 0.01$.

\section{Radiation from Turbulent Flares}\label{sect:flares}

\begin{figure}
\centering
    \includegraphics[trim={0cm 0.0cm 0cm 0.0cm}, clip=true, width=0.46\textwidth]{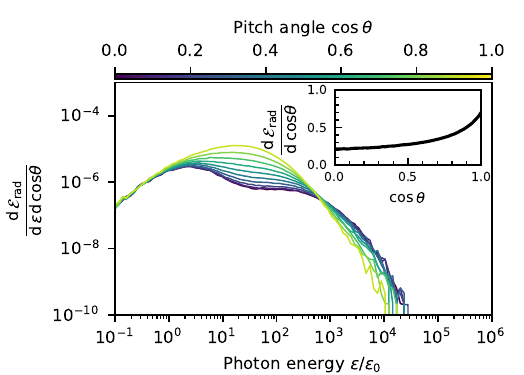}
    \caption{\label{fig:rad_spec_weak}
Spectrum of the time-integrated radiation from the fiducial 2D model of a turbulent flare with $\sigma_0 \approx 16$ and cooling parameter $\A \approx 0.003$. 
Curves with different colors show the spectra emitted at different angles $\theta$ with respect to the background magnetic field $\vec{B}_0$ (color code is indicated at the top). 
Inset in the top-right corner shows the angular distribution of the total energy output ${\cal E}_{\rm rad}$, normalized to the initial turbulence energy ${\cal E}_F^0$.
    }
\end{figure}

\begin{figure}
\centering
    \includegraphics[trim={0cm 0.0cm 0cm 0.0cm}, clip=true, width=0.46\textwidth]{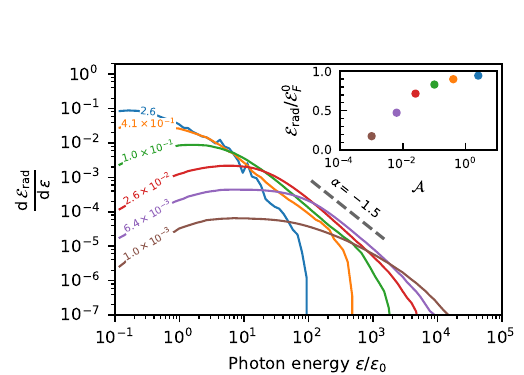}
    \caption{\label{fig:rad_specs_s10}
    Time-integrated radiation spectra calculated for models with 
$\A \approx $
$\Ten{1.0}{-3}$ (brown), 
$\Ten{6.4}{-3}$ (purple),
$\Ten{2.6}{-2}$ (red),
$0.10$ (green),
$0.41$ (orange), 
and $2.6$ (blue). 
Inset in the top-right corner shows the radiative efficiency $\eneR / \eneFz$, as a function of $\mathcal{A}$.
Dashed line indicates a characteristic spectral index of $\alpha = -1.5$.
}
\end{figure}

Our simulations follow radiative losses individually for each particle and allow us to study the produced radiation. 
Below we show the results for a simple emission model where the energy is passed to radiation via Thomson scattering of isotropic background soft photons with initial energies $\epsilon_0$. 
On average, a particle with Lorentz factor $\gamma$ upscatters the background photons to energy \citep{Rybicki_1985}
\begin{equation} \label{eq:scat}
\epsilon=\frac{4}{3} \beta \gamma^2 \epsilon_0.
\end{equation}
In the Thomson scattering regime, $\epsilon$ is a small fraction of particle kinetic energy $(\gamma-1)m_ec^2$, and so the particle is cooled through a large number of scattering events.
For simplicity, we assume that each scattered photon gains the average energy $\epsilon$ given by Equation~(\ref{eq:scat}).
We also assume that the upscattered photon is emitted into the direction of the particle velocity $\vec{\beta}$.
This gives a good approximation to the overall spectral and angular distributions of the produced radiation, especially when the $e^\pm$ plasma is relativistic. 
We calculate the emitted radiation by using a sample of $10^7$ particles, whose histories are followed throughout the simulation of the turbulent flare up to time $t=20\,l_0/c$.

Figure~\ref{fig:rad_spec_weak} shows the time-integrated radiation spectrum found in our fiducial 2D flare model in the weak-cooling regime, $\A \approx 0.003$.
The particle distribution during the flare extends from $\gamma\sim 1$ to $\gc\sim \A^{-1}\approx 100$, and, correspondingly, the radiation spectrum extends from $\epsilon\sim\epsilon_0$ to $\epsilon\sim \A^{-2}\epsilon_0$.
The emitted spectrum is shown for different viewing angles $\theta$ with respect to the background magnetic field $\vec{B}_0$, and one can see the strong anisotropy at $\epsilon\gg\epsilon_0$, where emission is dominated by nonthermal particles.

The observed anisotropy is shaped by the two different mechanisms of particle acceleration.
The impulsively accelerated particles in the reconnection layers emit mostly along $\vec{B}_0$, within an opening angle of $\theta \lesssim 30^\circ$ (this is a typical pitch angle of the particle when it escapes the reconnection layer).
The impulsive acceleration along $\vec{B}_0$ dominates the emission around the characteristic energy $\epsilon \sim \sigma_0^2 \epsilon_0$.
By contrast, the more energetic, stochastically accelerated, particles move preferentially at large pitch angles $\theta$. 
Their emission contributes to the spectrum at $\epsilon\gg \sigma_0^2 \epsilon_0$ and peaks at $\theta=\pi/2$.
Note that most of the flare energy is emitted by the impulsively accelerated particles in the current sheets, and therefore peaks at small $\theta$.

We have also calculated the produced radiation for the same flare model but in the strong-cooling regime, with $\A\gg\sigma_0^{-1}$. 
Then stochastic acceleration is suppressed, and therefore, the radiation spectrum is suppressed at $\epsilon\gg \sigma_0^2\epsilon_0$.
Impulsive acceleration in the current sheets remains efficient and continues to generate radiation beamed along $\vec{B}_0$ at pitch angles $\theta \lesssim 30^\circ$.

Figure~\ref{fig:rad_specs_s10} shows how the time-integrated radiation spectrum changes with $\A$. 
At low $\A$ the spectrum is very broad, with a weak contribution from the thermal plasma.
The position of the high-energy cutoff $\epsilon_c$ scales approximately as $\A^{-2}$, as long as $\A<\sigma_0^{-1}$. 
In the strong-cooling regime, there is an increasingly prominent emission from the thermal plasma, and the high-energy tail is limited to $\epsilon\lesssim \sigma_0^2\epsilon_0$.

The inset in Figure~\ref{fig:rad_specs_s10} shows the radiative efficiency of the flare ${\cal E}_{\rm rad}/{\cal E}_{F}^0$ (measured at $t=20\,l_0/c$) as a function $\A$. 
One can see that ${\cal E}_{\rm rad}/{\cal E}_{F}^0$ approaches unity at $\A\gtrsim 0.1$. 
In the weak-cooling regime, the efficiency varies approximately as $\ln\A$.

The characteristic luminosity of the flare may be estimated as ${\cal E}_{\rm rad}/t$, where $t\sim 10\,l_0/c$ is the characteristic duration of the main dissipation phase. 
In the radiatively efficient regime, the emitted luminosity will approximately track the overall dissipation rate, which decays approximately as $t^{-1}$. 
The strong intermittency of dissipation also implies that the radiation output is variable on
a broad range of timescales, from $\sim l_0/c$ to the kinetic tearing scale $l_c/c \approx (\sqrt{\sigma} \omp)^{-1}$ \citep[see also][]{Zhdankin_2020}.
Since the observed luminosity is integrated over the entire flare region, its temporal variations should be the strongest on the macroscopic timescales of $c/l_0$.

\section{Conclusions}\label{sect:conclusions}

We have studied decaying relativistic turbulent flares in a magnetically dominated collisionless pair plasma by means of kinetic PIC simulations.
The simulations were preformed with the open-source code \textsc{Runko} \citep{Nattila_2019}.
The turbulent flare was initiated by a macroscopic perturbation of the magnetic field on a large injection scale $l_0$.
This perturbation immediately led to the formation of large-scale current sheets between colliding magnetic flux ropes, and developed motions on a broad range of scales down to the microscopic plasma scale.

We observed heating of the thermal plasma and acceleration of nonthermal particles, in agreement with earlier simulations \citep{Zhdankin_2018,Comisso_2019}. 
The acceleration proceeds in two stages.
First, particles are almost impulsively accelerated to the Lorentz factor $\gamma_{\mathrm{inj}} \sim \sigma_0$ by a nonideal field $E_{\parallel}$  inside current sheets.
The impulsive acceleration is followed by stochastic acceleration to $\gamma \gg \gamma_{\mathrm{inj}}$ by the ideal MHD field $\boldsymbol{E}_\perp=-\boldsymbol{v}\times\boldsymbol{B}/c$ induced by the turbulent bulk motions.
In the absence of radiative cooling, this mechanism is capable of accelerating particles up to the maximum Lorentz factor $\gamma_0$ at which the particle Larmor radius becomes comparable to the driving scale $l_0$.
In most of our simulations, the plasma magnetization is $\sigma_0\approx 16$ and $\gamma_0\approx 500$. 
This gives the regime of $\gamma_0\gg\sigma_0$, as expected around compact objects.

Particles in turbulent flares near compact objects are subject to radiative cooling. 
In this paper, we studied how this changes the energy release process.
The strength of radiative cooling in our simulations is described by the dimensionless parameter $\A$ (Equation~\ref{eq:A}), which is similar to the usual compactness parameter \citep[e.g.,][]{Guilbert_1983}. 
It approximately equals the ratio of the light-crossing timescale $l_0/c$ to the cooling time $\tc(\gamma)$ of particles with $\gamma\approx 1$. 

Our results are as follows:
\begin{enumerate}
\item 
Driving of magnetically dominated plasma with strong perpendicular $\delta B \sim B_0$ perturbations leads to formation of large-scale magnetic flux ropes.
The flux ropes develop mildly relativistic shearing motions, as a consequence of the unbalanced initial state.
These motions shape the evolution of the decaying turbulent flare.
Magnetic reconnection between the colliding and merging ropes leads to intermittent plasma heating and nonthermal particle acceleration.
Plasmoids of various scales are ejected from the reconnection layers; 
however, large-scale shear motions prevent formation of long plasmoid chains.

\item
Radiative losses enhance the contrast of density variations in the turbulent plasma, although the spectrum of magnetic field fluctuations is not changed for all $\A<3$ studied in this paper. 
Radiative losses affect the process of energy deposition into the plasma when $\A>\gamma_0^{-1}$.
\item 
We observed two radiative regimes, which we called \textit{weak cooling} ($\A<\sigma_0^{-1})$ and \textit{strong cooling} ($\A>\sigma_0^{-1}$). 
Stochastic acceleration is suppressed in the strong-cooling regime; 
then only impulsive acceleration in current sheets generates nonthermal particles (with Lorentz factors $\gamma\sim \ginj\sim \sigma_0$). 
Note that a similar transition occurs in radiative magnetic reconnection \citep{Beloborodov_2017, Werner_2019, Sironi_2020}.

\item
In the entire range of $\A$ studied in this paper, the plasma sustains a well-defined  thermal component heated by turbulent reconnection.
It is formed by particles that never experienced an \textit{injection event}---a sudden energy gain in a current sheet. 
This criterion cleanly separates thermal and nonthermal components, even when the overall particle distribution appears to have a broad shape with no obvious separate components (see also \citealt{Comisso_2019}). 
The thermal component defined in this way is found to follow the Maxwell-J\"uttner distribution.

\item
The dissipated energy is distributed between the thermal and nonthermal particle populations with comparable rates as long as $\A\ll \sigma_0^{-1}$.
The partitioning of dissipated energy changes with increasing $\A$ as shown in Figure~\ref{fig:energy_budget}. When $\A$ reaches $\sim 1$, practically all dissipated energy is given to (and radiated by) the low-energy thermal particles.

\item 
Radiative losses offer a new tool to analyze the process of particle acceleration in PIC simulations.
In the weak-cooling regime, the losses compete with stochastic acceleration and impose a \textit{ceiling} $\gc$.
Its value depends on the diffusion coefficient in the energy space $D_\gamma$ that drives acceleration.
Our simulations demonstrate that there is no universal $D_\gamma$. 
Analysis of small sub-domains of the box shows that the global spectrum $f(\gamma)\propto\gamma^{-3}$ is the superposition of hard local spectra with different cutoffs $\gc$ \citep[see also][]{Lemoine_2020}.
We conclude that the total $f(\gamma)$ is shaped by the spatial and temporal intermittency of turbulent reconnection, which produces large variations of $D_\gamma$ across the box. 

\item
The simulations give the spectrum of the IC radiation emitted by the turbulent flare. 
We calculated the IC emission assuming an isotropic background of soft photons with energies $\epsilon_0$.
In the weak-cooling regime, the resulting spectrum is very broad, and its high-energy cutoff scales approximately as $\propto \A^{-2}$.
In the strong-cooling regime, the emission becomes more and more dominated by the radiation from the thermal plasma and the cutoff is located at $\epsilon \sim \sigma_0^2 \epsilon_0$.
The emission is anisotropic, and its angular distribution changes across the spectrum. 
The part of the spectrum dominated by current sheets, $\epsilon\sim \sigma_0^2\epsilon$, is beamed along the background magnetic field $\boldsymbol{B}_0$ within a characteristic opening angle $\theta\sim 30^\circ$.
The emission from stochastically accelerated particles dominates at energies $\sigma_0^2\ll\epsilon/\epsilon_0<\A^{-2}$ and peaks at $\theta=90^\circ$.

\item
Full 3D simulations of the turbulent flares show qualitatively the same behavior as the 2D simulations.
In particular, the magnetic spectrum, reconnection in the current sheets, and stochastic particle acceleration are all captured by the 2D model (see also \citealt{Comisso_2019}).

\end{enumerate}

In our simulations, the macroscopic driving scale is much greater than the plasma skin depth, $l_0 c/\omp > 100$, which corresponds to the maximum Lorentz factor achieved by stochastic acceleration $\gamma_0\sim 500$.
In real astrophysical objects, $l_0\omega_p/c$ can be larger by many orders of magnitude, which leads to enormous $\gamma_0$. 
This fact makes it important to know the precise scaling of the diffusion coefficient $D_\gamma\propto \gamma^\psi$.
We showed in Section~\ref{regimes} that $\gc/\ginj\sim \sigma_0^{-1}\A^{1/(3-\psi)}\gamma_0^{(2-\psi)/(3-\psi)}$ (Equation~\ref{eq:dissmeasure1}) and that a slight reduction of $\psi$ below $2$ enables stochastic particle acceleration even in the presence of strong cooling. 
This fact can have a strong impact on the observational appearance of turbulent flares in compact objects.

Another important aspect of turbulent flares is their spatial and temporal intermittency. 
It implies intermittency in the particle acceleration process and should leave imprints on the observed temporal structure of the produced radiation on timescales shorter than $l_0/c$.

\section*{Acknowledgments}
We would like to thank Luca Comisso, Lorenzo Sironi, Emanuele Sobacchi, and Daniel Gro\v{s}elj for helpful discussions, 
\AB{and the referee for useful comments that helped improve the paper.} 
The simulations were performed using resources provided by the Swedish National Infrastructure for Computing (SNIC) at PDC and HPC2N.
A.M.B. is supported by NASA grant NNX~17AK37G, NSF grants AST~1816484 and AST~2009453, Simons Foundation grant \#446228, and the Humboldt Foundation.

\bibliographystyle{aasjournal}
%\bibliography{refs}

\end{document}